\documentclass[prx,showpacs,showkeys,twocolumn,10pt]{revtex4-1}
\usepackage{amsfonts,bbold,amsmath,amssymb,graphicx,epstopdf,verbatim,dsfont,color}
\usepackage[english]{babel}
\newcommand{\kk}{\mathbf{k}}
\newcommand{\xx}{\mathbf{x}}
\newcommand{\pp}{\mathbf{p}}

\newcommand{\vv}{\mathbf{v}}

\newcommand{\tr}{{\rm Tr}}

\begin{document}

\title{Quantum statistical gravity: time dilation due to local information in many-body quantum systems}
\author{Dries Sels}
\affiliation{TQC, Universiteit Antwerpen, Universiteitsplein 1, B-2610 Antwerpen, Belgium}
\affiliation{Department of Physics, Boston University, Boston, MA 02215, USA}
\author{Michiel Wouters}
\affiliation{TQC, Universiteit Antwerpen, Universiteitsplein 1, B-2610 Antwerpen, Belgium}
\date{\today}

\begin{abstract}
 We propose a generic mechanism for the emergence of a gravitational potential that acts on all classical objects in a quantum system. Our conjecture is based on the analysis of mutual information in many-body quantum systems. Since measurements in quantum systems affect the surroundings through entanglement, a measurement at one position reduces the entropy in its neighbourhood. This reduction in entropy can be described by a local temperature, that is directly related to the gravitational potential. A crucial ingredient in our argument is that ideal classical mechanical motion occurs at constant probability. This definition is motivated by the analysis of entropic forces in classical systems, which can be formally rewritten in terms of a gravitational potential.
 
\end{abstract}
\maketitle

\section{Introduction}

After almost one century of coexistence, the relation between Einstein's theory of general relativity and quantum physics is still not well understood. At the same time, the precise relation between microscopic quantum physics and macroscopic classical physics has not been completely demystified. There are some suggestions that these two problems are related \cite{penrose14,tegmark,joos2013,gell-mann1993,hu2014}. 

Two fundamentally different strategies are used to relate the quantum to the classical. The first one is based on the wave-particle duality and is most succinctly expressed in the path integral formulation. The correspondence between the classical and the quantum is here mathematically very direct, because it is the same action that appears in both theories. It can therefore be used to go from the classical to the quantum and vice versa. A familiar example is electromagnetism: Maxwell's equations are derived from the quantummechanical path integral by means of the stationary phase approximation.

The second strategy, statistical physics, is fundamentally different. From a quantum theory, thermodynamic relations can be computed, but those thermodynamic relations cannot be quantised. Still, it is the only method suited for the description of complex macroscopic systems, about which we only have thermodynamic and hydrodynamic information \cite{zubarev,grandy2012}.

The main efforts to find a quantum mechanical description of gravity have been based on the first method \cite{polchinski,rovelli} but also the second strategy has been explored \cite{jacobson1995,Padmanabhan2005,verlinde2011}. The latter efforts go under the name of thermodynamic or entropic gravity. Conceptually, this strategy seems preferable, since the physics of gravity deals with macroscopic objects, that have nonzero entropy and that are coupled to environments \cite{zurek2003,wigner1995,schlosshauer2007}. In thermodynamic gravity, the gravitational interaction is seen as emergent rather than as an explicit ingredient in the microscopic theory. Gravity being the most universal force in the universe, it would ideally emerge in any complex quantum theory. In this paper, we will argue that this might be the case. 

The first ingredient in our argument is that the presence of matter at some place in the universe constitutes information, defined as missing entropy \cite{schrodinger1944,brillouin1953}. Within the framework of quantum mechanics, when knowledge about a particular realisation of the system is available, the incompatible part of the wave function has to be projected out.
When the quantum system has entanglement, a local projection also influences the probability distribution in its vicinity. 
We will show that the effect of local information on its surroundings can be described by a position dependent `entanglement' temperature \cite{wong2013,kumar2015}, defined by approximating the local reduced density matrix by a Gibbs state. We then demonstrate that the inhomogeneity of the entanglement temperature is reflected in a spatial variation of the magnitude of the energy fluctuations.

\begin{figure}[t]
\includegraphics[width=0.6\columnwidth]{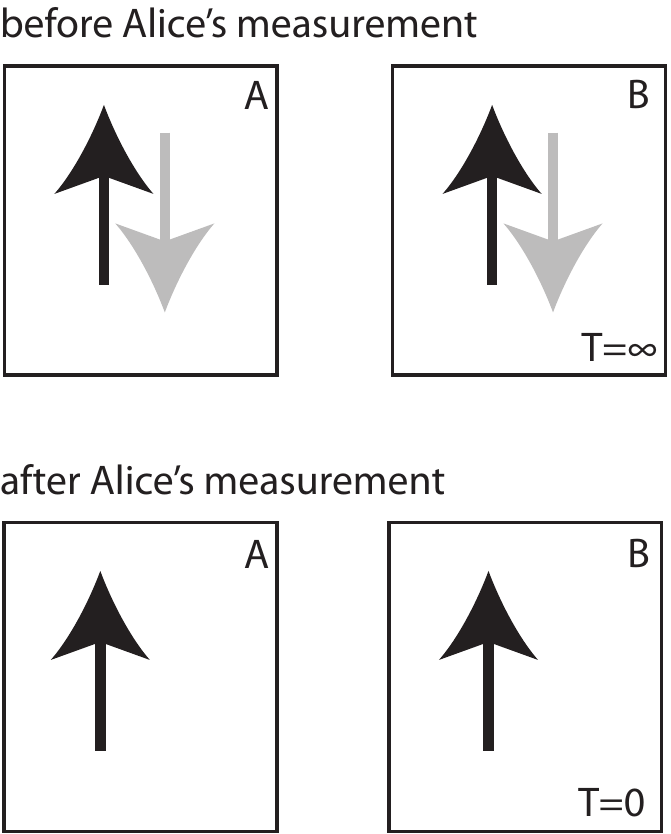}
\caption{A thermodynamic interpretation of the EPR experiment. Before the measurement, Bob's state is maximally mixed, corresponding to infinite temperature. Alice's measurement reduces Bob's state to a pure state, which is a zero temperature state.}
\label{fig:epr}
\end{figure}

This mechanism is most easily illustrated in the EPR setting. When the first qubit of a Bell state $| \psi \rangle = \frac{1}{\sqrt{2}}( \left |\uparrow \uparrow \right\rangle + \left|\downarrow \downarrow \right\rangle )$ is not measured, the reduced density matrix of the second qubit is maximally mixed, equivalent to infinite temperature. On the other hand, when the first one is measured in the $\uparrow,\downarrow$ basis, the second qubit is in a pure state, which corresponds to a zero temperature density matrix (see Fig. \ref{fig:epr}). 


It is actually well known that entanglement allows to make the connection between pure quantum states and statistical density matrices \cite{popescu2006,lloyd2013}. Tracing over environment degrees of freedom leaves the system typically in a canonical Gibbs state, even if the composite state is pure \cite{Goldstein}. We make a natural extension of this work by considering what happens if the environment is not fully traced over. 

The main assumption in our quantum statistical approach to gravity is that the magnitude of energy fluctuations in a region sets the energy scale for the local physics. This follows from the fundamental characterisation of phenomena by their probability. We will demonstrate that it is intimately related to the equivalence principle. 
`Gravitational red shifts' then appear because the entanglement temperature is lower closer to a source of information. We will also show that a position dependent entanglement temperature, implies the acceleration of semi-classical wave packets.

In our scenario, gravitational forces emerge as an interplay between quantum measurement and the statistical meaning of classical mechanics. Usually, we do not think about mechanical systems in terms of probability distributions, except in systems where `entropic' forces occur \cite{onsager,asakura}. We therefore start our paper with a short description of entropic forces in Sec. \ref{sec:entr}. It is illustrated that the entropic attraction between two objects is due to correlations in their joint probability distribution. We emphasise that entropic forces are not due to an increase in the entropy, but rather due to the constraint of motion at constant entropy. We point out the close analogy with Born-Oppenheimer forces.

In a quantum  system, the role of the probability distribution of local states is played by the reduced density matrix. Correlations between regions are reflected in the quantum mutual information \cite{vedral2002,Wolf}. An entropy and a temperature can be associated to the reduced density matrix. We show that local measurements can affect the temperature of the regions that are correlated with it. Our general arguments about local information in quantum systems are illustrated by calculations on a one-dimensional non-interacting fermion model in Sec. \ref{sec:Tloc}.

From the requirement of probability conservation of a wave packet, we show in Sec. \ref{sec:acc} that a temperature gradient leads to acceleration. The gravitational redshifts are discussed in Sec. \ref{sec:redshift}. By considering a global thermal equilibrium state, we recover in Sec. \ref{sec:tolman} the Tolman law, that describes how the temperature of a thermal equilibrium state varies in the presence of a gravitational field \cite{tolman}.

After some considerations on the relation between information and entropy (information is missing entropy \cite{brillouin1953}), we speculate on the nature of black holes in Sec. \ref{sec:BH}. In the logic of gravity emerging from information, we are led to conclude that black holes most naturally are the objects with maximal information, hence minimal entropy. We relate the origin of the Hawking-Unruh temperature \cite{hawking,unruh} to quantum fluctuations in the position of particles.

Most of the ideas that we use, have appeared in some form in other works. The last part of our paper is therefore devoted to the connections between our viewpoint and the related literature.

\section{Classical entropic forces \label{sec:entr}}

Theoretical classical mechanics deals with isolated systems, that satisfy energy conservation. This excludes systems that are coupled to baths and therefore it is only an approximate description of real physical systems, where friction due to coupling with an environment is always present.
In addition to friction, environments can also exercise forces on mechanical systems. These are known as entropic forces \cite{asakura} and are essential in e.g. depletion forces between colloidal particles in fluids, osmotic pressure and the elasticity of polymers. 

\begin{figure}[t]
\includegraphics[width=\columnwidth]{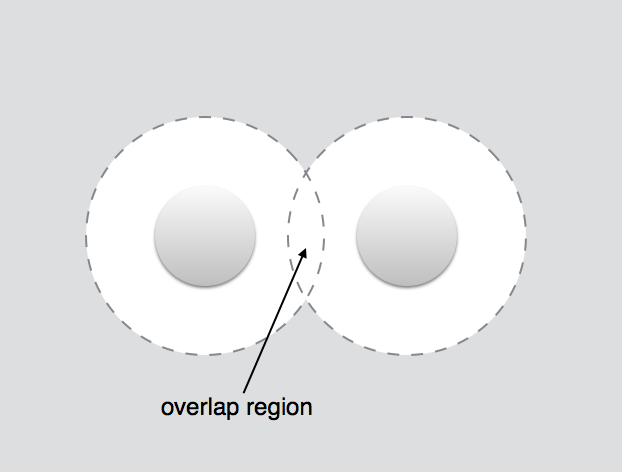}
\caption{Illustration of the entropically driven depletion force between two particles immersed in a fluid. When the two depletion regions overlap (marked by the dashed circles), the liquid has more volume and hence a larger entropy. This results in the attraction between the two immersed particles.}
\label{fig:depletion}
\end{figure} 

Let us start with an elementary analysis of entropic forces. It is the simplest setting in which one can see that probability conservation is a generalisation of energy conservation. We consider the classic example of two spheres immersed in a fluid (see Fig. \ref{fig:depletion}). Due to repulsive interactions between the sphere and the fluid, there is a depletion region around each sphere. The presence of the spheres therefore lowers the entropy of the fluid (it reduces the available volume for each molecule). When the depletion regions of the two spheres overlap, the entropy of the fluid increases. The probability in terms of their positions $\xx_{1,2}$ and momenta $\pp_{1,2}$ reads
\begin{equation}
p = \frac{1}{Z}e^{S(\xx_1-\xx_2)-\frac{p_1^2}{2m_1T}-\frac{p_2^2}{2m_2T}},
\label{eq:p_entr}
\end{equation}
where $T$ is the temperature (we set $k_B=1$) and $Z$ a normalisation constant. In the absence of the entropic term $S$ in the probability \eqref{eq:p_entr}, energy conservation is equivalent to probability conservation and the temperature $T$ is irrelevant. 

In the presence of $S$ on the other hand, probability and energy conservation are no longer the same. Energy conservation remains unaltered, but probability conservation requires
\begin{equation}
-T \ln p=\frac{p_1^2}{2m_1}+\frac{p_2^2}{2m_2}-TS(\xx_1-\xx_2) 
\label{eq:probcons}
\end{equation}
to be constant. A term $-TS$ is added to the energy, which gives rise to `entropic forces'
\begin{equation}
\frac{d\pp_i}{dt} = T \nabla_{\xx_i} S.
\label{eq:Fent}
\end{equation}

Such forces have been conclusively observed experimentally \cite{entropic_AFM}. We thus conclude that probability conservation is more powerful than energy conservation in  mechanical systems.

We wish to point out that phenomena that involve changes of the probability are considered to be outside of the scope of ideal classical mechanics, as follows from Liouville's theorem. In physical terms, it corresponds to the absence of dissipation. In `good' mechanical systems, the entropic term is negligible as compared to the other terms in \eqref{eq:probcons}, but as we will argue below, this omitted term could be responsible for the emergence of gravity when analysing classical objects in quantum systems.

The reason why the temperature never appears in classical mechanics is that one implicitly requires that the mechanical energy $E$ is much larger than the thermal energy. This means that classical mechanics is only concerned with statistically unlikely events, with $p \sim e^{-E/T} \ll 1$. This may seem contradictory to the standard classical to quantum correspondence, where the classical paths are the most likely, but corresponds to classical objects carrying a large amount of information (see Sec. \ref{sec:BH}). The meaning of the deterministic motion is that, conditional on being at position $x$ at time $t$, there is unit probability to be at a place $x'$ at time $t'$.

In quantum physics, where probabilities are the most elementary quantities, it is then natural to elevate these implicit aspects to the definition of a classical mechanical object: it is an unlikely excitation that evolves at constant probability.

The above discussion assumed a canonical picture where the temperature $T$ is fixed. To further illustrate the physics behind entropy driven forces, it is instructive to analyse the microcanonical situation, where the total energy is fixed.
When the spheres are accelerated, their kinetic energy is provided by reducing the internal energy of the molecules.
We will restrict our discussion to the case where one of the two spheres is fixed and the second sphere can move only in one dimension.
The discussion that follows below is generally valid for the motion at constant entropy. One could for example also apply it to the adiabatic expansion of a piston filled with a gas.

The condition of no entropy production $dS=0$ yields
\begin{equation}
dS = \frac{\partial S}{\partial x} d x + \frac{\partial S}{\partial E} dE =0,
\end{equation}
where $E$ refers to the energy of the gas. We reobtain the entropic force from Eq. \eqref{eq:Fent}
\begin{equation}
\frac{dP}{dt}=- \frac{dE}{dx} = T \frac{\partial S}{\partial x},
\label{eq:Fent2}
\end{equation}
where we have used the thermodynamic definition $T=\frac{dE}{dS}$ and denote the momentum of the sphere by $P$. The gas loses energy by performing work against the sphere, in order to keep the entropy constant.

In a Born-Oppenheimer language, the energy in the fast degrees of freedom (the gas) decreases under a displacement of the slow degree of freedom (the sphere). From the requirement of constant entropy for the gas, we can thus derive the Born-Oppenheimer potential, that governs the quantum-mechanical dynamics of the sphere. Note that it is not the increase of the entropy but its conservation that is responsible for the acceleration. Entropy increase corresponds to dissipation and hence falls outside of the scope of mechanical motion.

After the displacement, the decrease in internal energy is reflected in a change of the temperature. From equipartition, we have that $E=\frac 32 NT$, where $N$ is the number of molecules in the gas. This gives the relation
\begin{equation}
\frac{dE}{dx} = E \frac{d \ln T}{dx}.
\end{equation}
The momentum change can be rewritten in terms of the position dependence of the temperature as
\begin{equation}
\frac{d P}{d t}  = - E \frac{d \ln T}{dx}.
\label{eq:dP1}
\end{equation}
We conclude that a temperature gradient leads to a force that is proportional to the internal energy. 

Written in this form, the momentum change under adiabatic expansion of a gas is formally very close to the change in momentum due to gravitational forces. It can be written as
\begin{equation}
\frac{d P}{d t} =  -  \frac{E}{c^2}  \frac{d \phi}{dx},
\label{eq:dP2}
\end{equation}
where $\phi$ is the `gravitational' potential defined as
\begin{equation}    
\phi =c^2 \ln T
\label{eq:phiT}
\end{equation}

For classical systems, these formal manipulations do not yield any new physics beyond the standard thermodynamic analysis, but we will show below that measurements in quantum systems lead to a position dependence of the local temperature. Thanks to the analogy between temperature gradients and gravitational interactions, we are then led to the appearance of gravity as a consequence of local information in quantum systems.

\section{Local information and temperature in quantum systems \label{sec:Tloc}}

After these preliminary remarks on classical mechanics, we turn our attention to quantum systems. The central object in quantum mechanics of closed systems is its wave function. 
We will consider quantum systems that are defined on a lattice. A typical example is the Hubbard model. At first sight, one would have little hope for the emergence of relativity out of such systems, but Lieb and Robinson showed in their seminal work that causality emerges in such systems under very general conditions \cite{LR}. 

The time evolution of a quantum system is described by a unitary operator $U(t) = e^{- i \hat H t}$ (we set $\hbar=1$). A severe objection for this construction to be related to our universe is that it selects preferential coordinates. The point of our work is however that the parametric time $t$ does not necessarily correspond to physical time. In the regimes that we will discuss, the physical time depends on parametric time through a space-dependent time dilation factor, just as in the Newtonian limit of Einstein's gravity.

Before going to the technical analysis of a specific example, it is worth spending a few words on the issue of locality. While the natural mathematical space of quantum mechanics is the Hilbert space of its quantum states, physical systems are also endowed with a coordinate space. This coordinate space is important, because physical Hamiltonians are not arbitrary Hermitian operators acting on the Hilbert space, but show additional structure. In particular, the majority of physical Hamiltonians are local, which means that they couple only sites that are close to each other. It is for this class of Hamiltonians that Lieb-Robinson causality emerges. 

A second important property of the local Hamiltonians concerns the entanglement entropy of their ground states. If we only want to know local quantities, it is sufficient to know the reduced density matrix of the region we are interested in. Since this involves a loss of information, the reduced density matrix will have in general nonzero entropy, the entanglement entropy. The entanglement entropy is typically proportional to the number of sites in the local region. For ground states of local Hamiltonians however, it scales subextensively with system size \cite{Eisert}.

In order to make our discussion more concrete, we perform some calculations on a specific model. The choice of models is quite limited, since generic interacting quantum systems have a prohibitively large Hilbert space to perform explicit calculations. The simplest systems from a theoretical point of view are the ones with quadratic Hamiltonians, that describe free quasi-particles.
We will here consider the ground state of the fermionic hopping Hamiltonian of the form
\begin{equation}
\hat H =  -\sum_i \left[ \left( t\hat c_{i+1}^{\dagger} \hat c_i + h.c. \right) +\mu \hat c_{i}^{\dagger} \hat c_i) \right],
\label{eq:Hbog}
\end{equation}
where $t$ is the hopping amplitude and $\mu$ is the chemical potential. Apart from a Hamiltonian, we need to specify the quantum state of the system. Here, we will consider the ground state of the Hamiltonian, but the analysis could be extended to excited states as well. 

All expectation values within a subregion $A$ are described by the reduced density matrix $\rho_A = \tr_{S \setminus A} \rho$, where the trace is over all sites except the ones in A. Since $\hat H$ is a quadratic Hamiltonian, its ground state satisfies Wick's theorem. The reduced density matrix $\rho_A$ is therefore also specified by the second order correlation functions and is quadratic in the creation and annihilation operators on region $A$ as well:
\begin{equation}
\hat \rho_A = \exp(-\hat H_A).
\end{equation}
The `modular' or `entanglement' Hamiltonian can be written as
\begin{equation}
\hat H_A = \sum_{i,j} \hat a^\dag_i h_{ij} \hat a_j.
\end{equation}
The matrix $h_{ij}$ can be found by a diagonalisation of the correlation matrix (see Appendix for more information). 

\begin{figure}[t]
\includegraphics[width=\columnwidth]{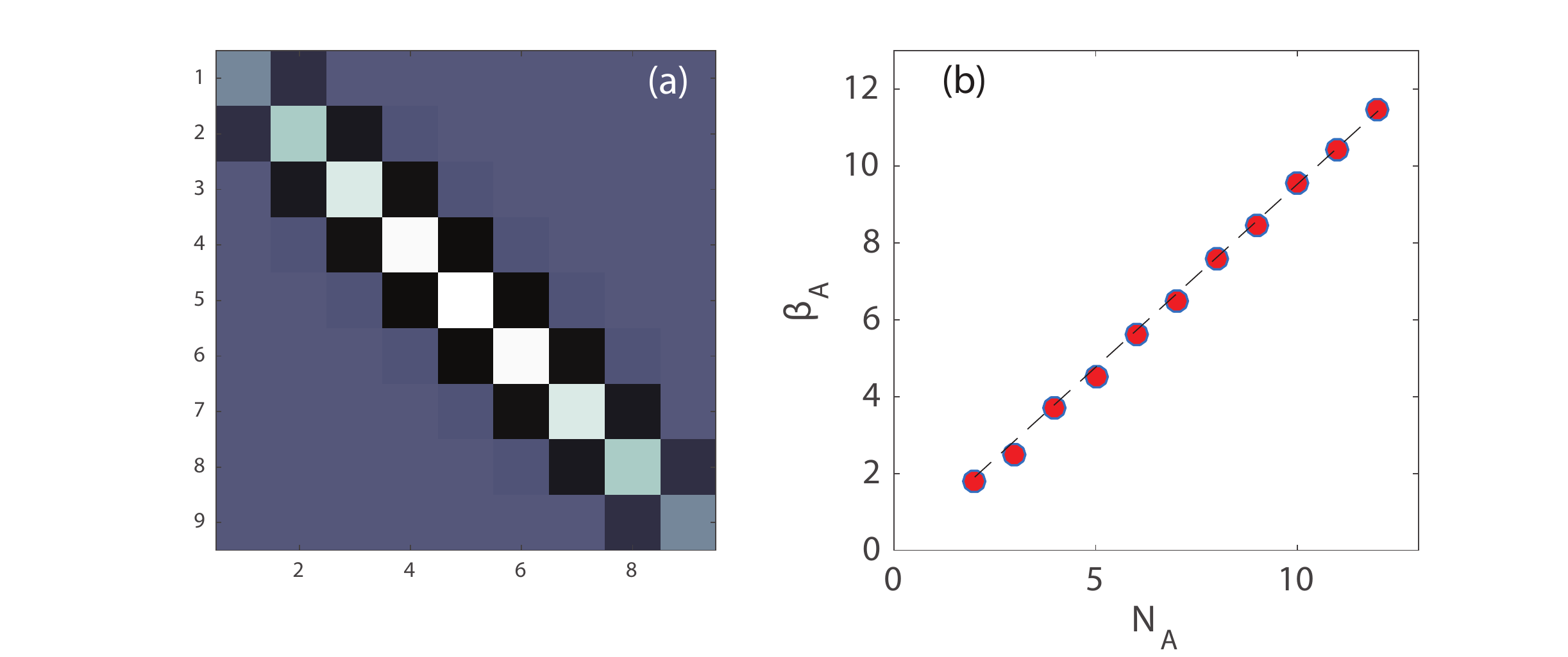}
\caption{(a) The entanglement Hamiltonian $h_{ij}$ (left) for the fermionic Hamiltonian \eqref{eq:Hbog} on a subregion of 10 sites. The structure of the original Hamiltonian is clearly visible, with hopping matrix elements next to the diagonal. (b) The entanglement temperature increases as a function of the system size.}
\label{fig:HM}
\end{figure}

An example for $N_A=10$ sites is shown in Fig. \ref{fig:HM}. The structure of the original Hamiltonian is clearly visible in the entanglement Hamiltonian. When comparing entanglement Hamiltonians for different subsystem sizes, one finds that the main effect is that the magnitude of the matrix elements increases. We thus find that the entanglement Hamiltonian is of the form $ h_A = \beta_A \tilde{h}_A $. We will call this scale dependent temperature $T_A$ the `entanglement' temperature \cite{wong2013,kumar2015}.
Fig. \ref{fig:HM}b shows the dependence of $\beta_A$ on system size, where $\tilde{h}_A$ is normalised by the largest tunneling matrix element.
A linear increase of the effective inverse temperature with system size is apparent, in agreement with analytical calculations exploiting the adS/CFT correspondence~\cite{bhattacharya2013}.

This behavior can be understood from the fact that, unlike the entropy of a Gibbs state, the entropy is not extensive. For 1-D free fermions it is well know that the ground state entanglement entropy is $S=1/3 \log(L)$. The entanglement temperature must thus decrease such that the entropy of a Gibbs state of the local (smooth) Hamiltonian equals the entanglement entropy.

Let us now consider a subregion $AB$ that consists of two disconnected parts $A$ and $B$, separated by a distance $R$. In terms of the entanglement Hamiltonian, there is a difference between $h_{AB}$ and the direct sum $h_A \oplus h_B$. Because more information is present in the compound subsystem $AB$, larger matrix elements are found in $h_{AB}$: the joint system is at a lower temperature than the individual systems. Physically, this means that the uncertainty about the state $A$ is reduced when information is obtained about system $B$.

It is instructive to look at the limiting cases $R=\infty$ and $R=0$ when $A$ and $B$ have the same number of sites. When $A$ and $B$ are infinitely far apart they should not be entangled and the total entanglement Hamiltonian is just the direct sum entanglement Hamiltonian, hence the entanglement temperature is the same as for the individual systems. However, when the two systems touch they simple form one system that is twice as big. It was already shown above that the entanglement temperature is only half of that of the separate subsystems.

\begin{figure}[t]
\includegraphics[width=\columnwidth]{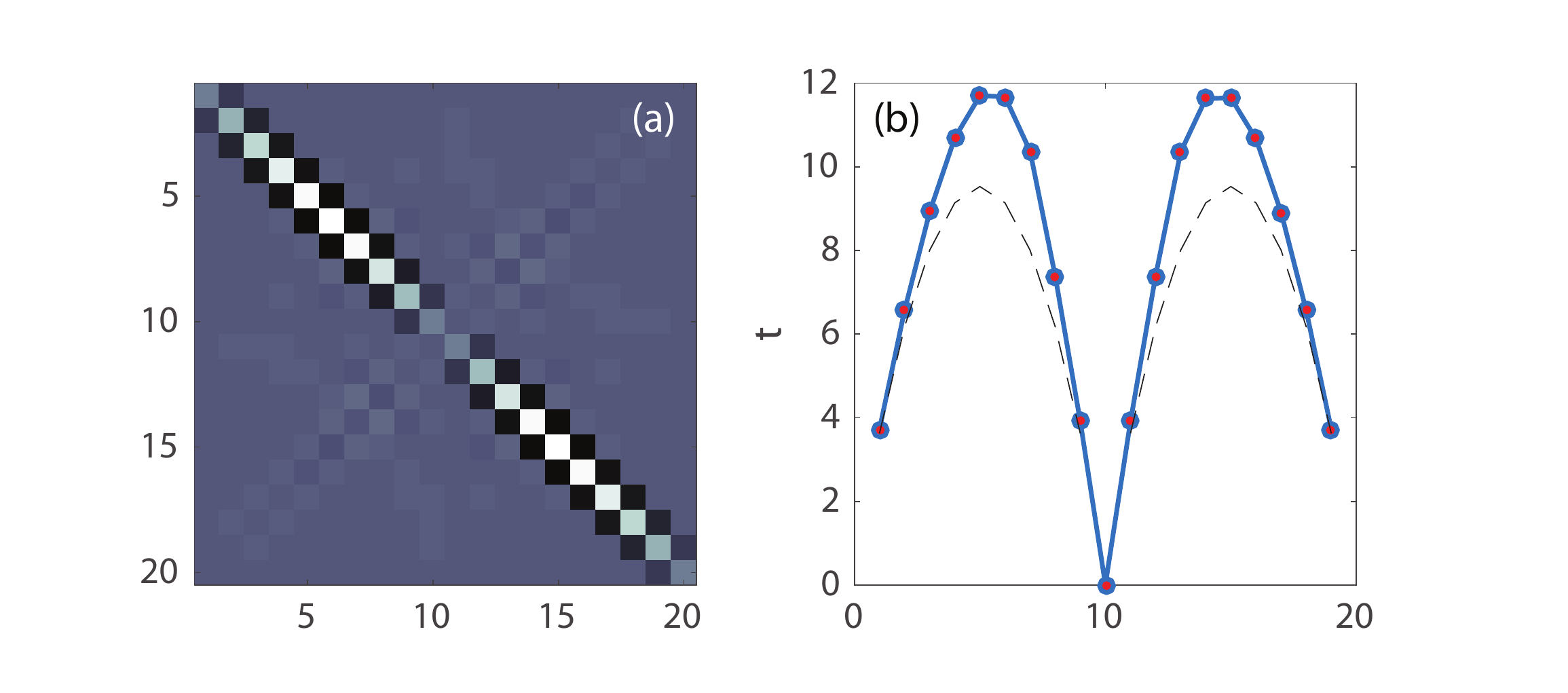}
\caption{(a) The entanglement Hamiltonian $h_{ij}$ for a subsystem consisting of two disjoint regions (10 sites each) separated by a distance of 15 sites. (b) Tunneling matrix elements (first off-diagonal) are compared with the case of a single region of 10 sites (dashed lines). }
\label{fig:HMAB}
\end{figure}

This is illustrated in Fig. \ref{fig:HMAB}, where we show the entanglement Hamiltonian consisting of two subregions of 10 sites that are separated by 15 sites. The left hand panel shows the full matrix, where one can clearly identify the direct sum of uncoupled entanglement Hamiltonians and some off-diagonal couplings.
The right hand panels show the tunneling matrix elements and compares them to the case of a single subregion (dashed lines). Two features stand out. First, the tunneling matrix elements are larger, corresponding to a larger inverse temperature: $\beta_{AB} > \beta_A$. Secondly, there is also an asymmetry in the matrix elements, which could be interpreted as a temperature gradient. 

\begin{figure}[t]
\includegraphics[width=\columnwidth]{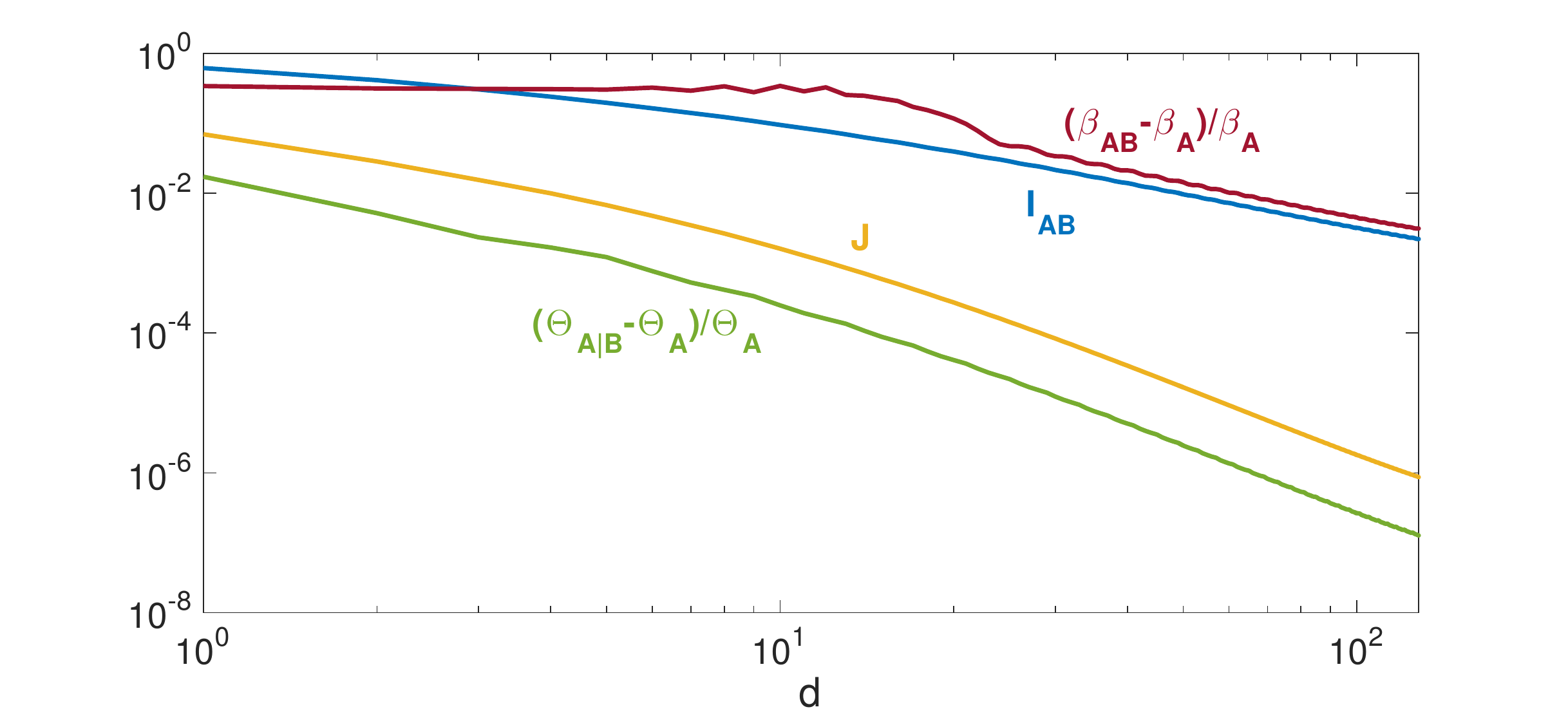}
\caption{The mutual information $I_{AB}$ as a function of the distance between two subregions  of sizes $N_A=N_B=10$ (blue), the effective temperature of the density matrix $\hat \rho_{AB}$ (red), the average conditional entropy $J$ for projection on the Schmidt basis in region $A$ (orange) and the energy average fluctuation width after projection on the Schmidt basis (green).}
\label{fig:IAB}
\end{figure}

The entropy of the total system $AB$ is also different from the sum of the entropies of $A$ and $B$. The difference between the two is the mutual information
\begin{equation}
I_{AB} = S_A + S_B - S_{AB},
\label{ref:IAB}
\end{equation}
which is always positive. Fig. \ref{fig:IAB} shows the mutual information as a function of the distance for our free fermion toy system (blue line). The mutual information is seen to decay slowly, with a power law behavior at large distance. This can be attributed to the fact that the system is gapless \cite{Eisert}. The red line shows the effective temperature $\beta_{AB}$ of the joint density matrix $\hat \rho_{AB}$, which shows the same behavior at large distances. This shows a clear connection between the increase in mutual information and the decrease in effective temperature when the two subregions are brought closer together.

For classical systems, one can rewrite the mutual information in terms of conditional entropies as:
\begin{equation}
I_{AB} = S(A) - \sum_{x_B} p(x_B) S(A|x_B),
\end{equation}
where the conditional entropy $S(A|x_B)$ is defined in terms of the conditional probability distribution as $S(A|x_B)= - \sum_{x_A} p(x_A|x_B) \ln p(x_A|x_B) $.

For quantum systems unfortunately, the situation is more complicated, because conditional probabilities are no longer simply defined in terms of joint probabilities, but by means of projection operators. It has been found that for quantum systems, the second definition of the mutual information is always smaller than the first one. This has led Zurek to introduce the notion of `quantum discord' as \cite{zurek2003discord}
\begin{equation}
D = I_{AB} - \max_{\Pi_j^A} J_{\Pi_j^A} (\rho),
\label{eq:discord}
\end{equation}
where $J$ is the conditional entropy that depends on the set of projective measurements $\Pi_j^A$.
Unfortunately, the optimisation problem over measurements has been proven to be NP complete \cite{huang}. 

If one is interested in the physics of a region $A$, one would want to know the reduced density matrix $\rho_A$. Imagine now that a measurement in region $B$ is performed. When there are correlations (quantified by the mutual information) between these two regions, the result of this measurement will affect the density matrix in region $A$. In general, the conditional entropy $S_{A|B}$ will be lower than $S_A$, with the mutual information as an upper bound on the average entropy reduction for a certain type of measurement. In view of the relation between entropy and temperature, this means that the temperature of the region $A$ depends on information about region $B$. In other words, a measurement in $B$ results in a position dependent temperature in its surroundings. Regions close to $B$ will be more affected than regions farther away, as indicated by the behavior of the mutual information.

The simplest class of projections to be performed on this system is a projection on the Schmidt basis in region $B$. While this is clearly not the best basis in the sense of Eq. \eqref{eq:discord}, it gives at least a lower bound on the average amount of information that can be gained with a projective measurement. From Fig. \eqref{fig:IAB}, it is seen that conditional entropy $J$ is lower than the mutual information and decays faster as a function of the distance, but it still shows a power law type decay.

\section{Energy statistics and mechanical acceleration \label{sec:acc}}

In order to make contact with fundamental physics, one should think about the density matrix before measurement as the `vacuum'. An unlikely measurement outcome would then correspond to the detection of matter. If matter is measured to be present in region $B$, this measurement then affects the statistics in region $A$, depending on the distance between $A$ and $B$.

The probability to have a fluctuation with energy $E$ and momentum $P$ in region $A$ can be written as the Fourier transform
\begin{equation}
p(E,P) = \int \frac{d\beta_E}{2\pi} \frac{d\beta_P}{2\pi} e^{i \beta_E E - i \beta_P \hat P} \langle e^{-i \beta_E (\hat H-\bar E_A) + i \beta_P \hat P} \rangle,
\end{equation}
where $\bar E_A$ is the average energy in region $A$. The average is taken with respect to the density matrix $\rho_A$: $\langle \hat O \rangle = \tr(\rho_A \hat O)$.

For large fluctuations, the saddle point approximation can be used. It allows us to write the probability distribution in terms of the Legendre transform of the characteristic function
\begin{equation}
\phi(\beta_E,\beta_P)= \ln  \langle e^{-\beta_E (\hat H-\bar E) +  \beta_P \hat P} \rangle.
\end{equation}
For the probability, we have
\begin{equation}
\ln p(E,P) = \min_{\beta_E,\beta_P} [\phi(\beta_E,\beta_P) + \beta_E E - \beta_P P ].
\end{equation}
We can rewrite this in a more physical way by introducing the velocity $v=\beta_P/\beta_E$ (and we set $\beta_E=\beta$). The probability then reads
\begin{equation}
\ln p(E,P) = \min_{\beta,v} [\phi(\beta,v) + \beta( E -v P )].
\end{equation}
This gives us the relations
\begin{equation}
E -vP=-\frac{\partial \phi}{\partial \beta}, \hspace{1cm} 
P = \frac{1}{\beta}  \frac{\partial \phi}{\partial v}.
\label{eq:EP}
\end{equation}

Let us now consider a situation where the probability distribution depends on the position $x$ of region $A$. As in the classical case, we can now obtain the entropic force by requiring constant probability for the fluctuation: 
\begin{equation}
d \ln p = d  [\phi(\beta,v,x) + \beta( E -v P )] = 0.
\end{equation}
Using the relations \eqref{eq:EP}, we can simplify this to
\begin{equation}
\beta d E - \beta v dP + \frac{\partial \phi}{\partial x} dx = 0.
\end{equation}
Since there is no external potential acting on the system, the magnitude of the energy fluctuation $E$ should remain the same. We then find with $v \frac{d}{dx} = \frac{d}{dt}$ that
\begin{equation}
\frac{dP}{dt} = \frac{1}{\beta} \frac{\partial \phi}{\partial x}.
\label{eq:dPdt}
\end{equation}

Let us now consider the situation with small velocity and neglect the momentum in the probability distribution. The typical distribution will be Gaussian in energy fluctuations, which corresponds to a Gaussian characteristic function
\begin{equation}
\phi(\beta) = \frac{\Theta^2(x)}{2} \beta^2.
\end{equation}
The Legendre transform reads
\begin{equation}
 E=-\Theta^2(x) \beta, \hspace{1cm} \ln p(E) = -\frac{E^2}{2\Theta^2(x)}.
\end{equation}
For the force \eqref{eq:dPdt}, we then get
\begin{equation}
\frac{1}{\beta}  \frac{\partial \phi}{\partial x}= \Theta(x) \Theta'(x) \beta = - E \frac{d \ln \Theta(x)}{dx},
\end{equation}
so that (for small velocities)
\begin{equation}
\frac{dP}{dt} =  - E \frac{d \ln \Theta(x)}{dx}.
\label{eq:dPentr}
\end{equation}
The change in momentum is proportional to the energy and the gradient of the standard deviation of the energy. This results in acceleration towards regions with less fluctuations. Note the analogy with the classical entropic force, where the internal energy is also reduced by the work performed against the spheres.

Alternatively, we could have derived the entropic force \eqref{eq:dPentr} by requiring a constant probability for the internal energy $E_{int}$
\begin{equation}
\ln p(E_{int},x) = - \frac{E^2_{int}}{2 \Theta^2(x)}.
\end{equation}
We then obtain immediately from 
\begin{equation}
d \left( \frac{ E_{int}}{\Theta(x)} \right) = 0 
\end{equation}
that there is a gradient in the internal energy, corresponding to a force
\begin{equation}
F= - \frac{d  E_{int}}{dx} =- E_{int} \frac{d \ln \Theta}{dx}.
\end{equation}
The principal ingredient that is responsible for the entropic force is the spatial variation of the energy fluctuations. As we have argued in the previous section, these can arise from  local information about the system. 
The requirement that the internal energy of classical objects follows the local temperature then implies a spatially dependent speed of time, i.e. a \emph{time dilation}. In the next section, we will further discuss how the local temperature is directly related to the red shift.

In Fig. \ref{fig:IAB}, we plot the energy variance after projection on the Schmidt basis (as for the computation of $J$) with a green line. The spatial dependence of the energy fluctuations follows the behavior of the conditional entropy, establishing a link between entropy and energy fluctuations. This suggests that the energy fluctuations $\Theta$ are proportional to the entanglement temperature. We will therefore call $\Theta$ the `local temperature' in the following.

If we interpret the energy $E$ in Eq. \eqref{eq:dPentr} as the relativistic rest energy $E=mc^2$ of the excitation in the sense that $p=m v$, we obtain the acceleration
\begin{equation}
a =- c^2  \frac{ d \ln \Theta}{d x}. 
\end{equation}
The local temperature can then be identified with the Newtonian potential as
\begin{equation}
\phi = c^2 \ln \Theta.
\label{eq:phi}
\end{equation}

Let us recapitulate how in our view gravity emerges from quantum mechanics
\begin{enumerate}
\item Local properties of quantum systems are described by a reduced density matrix, characterised by a local temperature.
\item Local information leads to a spatial dependent temperature.
\item In order to conserve their probability, excitations in regions with different temperatures have different energies, which leads to acceleration.
\end{enumerate}

Looking back at the ingredients that led to the (weak) equivalence principle, the most crucial step is the requirement that $E/\Theta(x)$ is constant, motivated by constant probability. This is our translation of the physics of ideal mechanical motion into the language of statistics.

Einstein translated the physics of ideal mechanical motion (free falling objects) into the language of classical field theory as `general covariance': the form of the equations should be independent of the coordinates. This principle has turned out to be hard to implement in a quantum setting. The reason could be the fact that quantum mechanics is fundamentally a statistical theory, requiring a direct formulation of physical processes in a probabilistic language.

\section{Gravitational redshifts \label{sec:redshift}}

Gravitational redshifts can also be seen as a direct consequence of the position-dependent temperature. When light that is emitted by a certain source (e.g. a burning candle) is detected, we infer its expected energy from the observed phenomenology. According to our line of reasoning, it is however more natural to take the dimensionless ratio of the energy to the local temperature $E/\Theta(x)$ to be the fundamental characterisation of a phenomenon. 
This argument is in line with standard thermodynamics where only relative temperatures can be measured (e.g. by means of the Carnot efficiency). The element that we add to this discussion is that quantum entanglement allows to define a fundamental temperature of empty space. Under our assumption this temperature sets the scale for the local physics, the same phenomenon at a different position can have a different energy.
We come again to time dilation as a consequence of local information.

When the light propagates, it does not change its frequency, because the propagation is governed by the hamiltonian $\hat H$ and there is no internal energy that can change with position. We then find the frequency of a photon emitted at position $x$ to differ from a photon emitted by the same phenomenon at position $x'$ by a factor
\begin{equation}
\frac{\nu(x)}{\nu(x')}=\frac{\Theta(x')}{\Theta(x)}
=e^{\Delta \phi/c^2},
\label{eq:tdil}
\end{equation}
where $\Delta \phi = \phi(x)-\phi(x')$. For the last equality, we have used the definition of the gravitational potential \eqref{eq:phi}. It gives the same relation between the red shift and the gravitational potential as in general relativity \cite{verlinde,wald}.

\section{The Tolman effect \label{sec:tolman}}

It is also interesting to consider a thermal equilibrium state of the real Hamiltonian $\hat H$ at nonzero temperature $T$, that is constant in space. According to the discussion of gravitational redshifts, this constant temperature will appear to be different at different positions. 
The physical temperature measured with a local thermometer then corresponds to the ratio of the actual temperature $T$ to the local temperature: $T_{\rm phys}(x)=T/\Theta(x)$.  

Using \eqref{eq:tdil}, we recover the same relation between the spatial dependence of the physical temperature and the red shift
\begin{equation}
\frac{T_{\rm phys}(x) }{T_{\rm phys}(x')} = \frac{\Theta(x')}{\Theta(x)} = \frac{\nu(x)}{\nu(x')}
\end{equation}
as in the Tolman law in a static metric~\cite{tolman}. 

\section{Speculations on black holes and the Hawking-Unruh temperature \label{sec:BH}}

The mysteries concerning black holes have played a big role in the quest for a quantum mechanical theory of gravity. Specifically, the understanding of their entropy and loss of information constitute important theoretical challenges. Before speculating on how black holes may fit into our picture of quantum statistical gravity, we start by making some general comments on the relation between information, probability and entropy.

When a measurement is performed on a quantum system with an a priori density matrix $\rho_0$, the acquired information changes the density matrix to $\rho$. 
The amount of information that is obtained by this measurement can be quantified by the relative entropy or the Kulback-Leibler distance \cite{vedral2002}
\begin{equation}
I(\rho)\equiv D_{KL}(\rho   || \rho_0) = -\tr(\rho \ln \rho_0) + \tr(\rho \ln \rho).
\label{eq:DKL}
\end{equation}
It expresses how surprising measurements are on a system with density matrix $\rho$, when you thought that the density matrix was $\rho_0$.

The simplest situation, which is the most relevant for typical thermodynamic systems, is when all $N_0$ states in $\rho_0$ can be assumed to have the same probability (microcanonical ensemble) and in $\rho$, a subset of $N$ states in $\rho_0$ (defining a `macrostate') are occupied. 
We then find for the information $I = S(\rho_0)-S(\rho)$, were $S(\rho_i)=\ln N_i$ is the entropy. In words, the information in a state is equal to the missing entropy \cite{brillouin1953}.
It is also immediate to relate the information to the probability, as first done by Einstein
\begin{equation}
p=e^S = e^{S_0 - I}.
\end{equation}
Here we see clearly that the most unlikely macrostates contain the largest amount of information. 

In the context of gravity, the macrostates are specified by a certain distribution of matter. But we should turn the argument around and try to define matter through the information in the local density matrix. This problem is however beyond the scope of the present discussion. We will restrict to some speculations, indicating how we see possibilities for the understanding of black hole physics from our statistical perspective.

When the amount of matter in a certain region is defined by the information in the density matrix, i.e. the missing entropy, the natural upper bound to the amount of matter a given region can contain is the entanglement entropy of the unmeasured quantum state. The remaining entropy of the maximum information state is then zero. The objects with maximal information (matter) in a given region of space are most naturally identified with black holes. It seems however unlikely that the entanglement entropy can be zero for stable objects, because the Hamiltonian couples all the regions in space, leading to decoherence, but it seems reasonable to conjecture that black holes are the stable objects with the minimal entropy that can be attained in a given region.

Conceptually, this would be very comforting: the elementary objects that appear in the culmination of classical physics, Einstein's theory of general relativity, would carry the minimal quantum uncertainty. The inaccessibility of the black hole interior then simply comes from the fact that its state is fixed by the requirement of minimal entropy. The explosion of a star at the end of its life is the ultimate cooling to a state with the least possible entropy. 

It is impossible to speculate on the quantum nature of black holes without mentioning the Hawking temperature, caused by quantum fluctuations \cite{hawking}. We believe that the mechanism could be quite simple. In our discussion of mechanical acceleration, we have seen that the spatially dependent temperature results in a spatially dependent internal energy. When there is an uncertainty in the position, this results in an uncertainty of the energy $\Delta E = (\partial_x E) \Delta x $. When we now take for the minimum uncertainty the `Compton' wave length $\Delta x=\hbar c/E$, we find $\Delta E = (\hbar c/E) \partial_x E $, which can be written as
\begin{equation}
\Delta E = \frac{\hbar a}{c}  \sim T_{U}.
\label{eq:TU}
\end{equation}
where we have used that the acceleration is equal to $a =(c^2/E) (d E/d x)$.
We obtain an uncertainty on the energy that is of the order of the Unruh temperature $T_U=\hbar a/(2\pi c)$. For a black hole, the acceleration is replaced by the surface gravity, resulting in the Hawking temperature $T_H$.

In our picture, there is no information paradox~\cite{hawking1976,preskill}, since it is the natural evolution of unitary quantum mechanics to wash out a local suppression in the entanglement entropy. It is the basic mechanism that leads to thermalisation in quantum many body systems. The evaporation of a black hole is then in principle no different to the mixing of hot and cold tea. The physical interpretation of this solution is however quite exotic: it is not a remnant, but the vacuum that is entangled with the emitted radiation. In our view, empty space has a higher entanglement entropy than space that contains matter excitations (information). This is opposite to the traditional assumption that entropy is carried by quasi-particles. 

\section{Lorentz transformations}

So far, we have discussed time dilation of the `gravitational' type. A more elementary type of time dilation already occurs in special relativity under Lorentz boosts. We want to discuss here how one can understand the need for this type of time dilation from an information analysis. 

A simple theoretical model in which Lorentz transformations can be investigated is that of a low temperature superfluid Bose gas. It has emergent Lorentz invariance for its excitations, that have a linear dispersion $\omega_\kk = c |\kk|$.
Let us consider a box of volume $V$ that contains thermal excitations at temperature $T$ and moves at a speed $v$ with respect to the superfluid. Its entropy equals \cite{volovik}
\begin{equation}
S = C \frac{T^3 V}{(\sqrt{1-v^2/c^2})^4},
\end{equation}
where $C$ is a numerical constant. This formula shows that the entropy depends on the velocity. In order to keep the entropy constant under a boost, the temperature should change. According to our previous arguments, a change of the temperature corresponds to a rescaling of time. When the lengths are measured with sound waves, the time dilation will have to be accompanied by a length contraction in order to keep the measured distances the same. To a rescaling of the temperature with a factor $T \rightarrow \gamma^{-1} T$ then corresponds a rescaling of the volume with the same factor $V \rightarrow \gamma^{-1} V$. 

The entropy of the moving system is then equal to the entropy at rest when $\gamma^{-1}=\sqrt{1-v^2/c^2}$. We thus recover the usual Lorentz transformation of the temperature four vector~\cite{volovik} $(\beta,0) \rightarrow (\gamma \beta, \gamma  \vv \beta)$ from the requirement that the entropy should be conserved under boosts.

Here, we used as an example the superfluid Bose gas with emergent Lorentz invariance, but it could expected that the Lorentz invariance of the entropy holds for all quantum many body systems that show causality in the sense of the Lieb-Robinson bound. Unfortunately, we are not aware of any proof of this statement.

\section{Relation to other works}

The intimate relation between temperature and information dates back to Maxwell's demon thought experiment \cite{leff2002,maxwell}. He discussed how a demon that had access to the velocities of individual atoms in a gas can reduce the temperature by opening a door when a slow atom passes. In other words, when we have information about the system, its effective temperature is lower. We add to this that information in a quantum system affects the temperature in its vicinity through entanglement.

The connection between measurements in quantum mechanics and time dilation was already made by Bohr in his legendary discussions with Einstein on the Heisenberg uncertainty principle \cite{bohr}. Einstein did not want to accept the probabilistic character of quantum mechanics, which Bohr saw to be the very foundation of the theory. When Einstein proposed to violate the energy-time uncertainty principle through a measurement of time with a clock and energy with a scale, Bohr saved it by invoking the uncertainty in the time measurement induced by the uncertainty in the position of the clock. As we have discussed in Sec. \ref{sec:BH}, this could be the underlying mechanism for the Hawking-Unruh effect.

As discussed in the introduction, the famous EPR paper on the incompleteness of quantum mechanics \cite{epr} is directly connected to our mechanism for gravity. The main conclusion of the paper was phrased as the incompleteness of quantum theory, because of the instantaneous effect of the projection on distant regions. It is important to discuss this objection to quantum mechanics in the context of our work. An obvious criticism to our mechanism would be that it allows gravitational interactions to be mediated faster than light. We believe that this problem should  be addressed by a deeper analysis of the `measurement problem'. 
In this paper, we have not addressed measurements in detail, due to a lack of mathematical techniques to properly deal with it. But when we say that there is information about a part of the system, we do not think of suddenly doing a measurement. We are aware of the fact that the earth exists, without us being able to decide that we are going to make a measurement. We rather think of measurement as `post-selection' than an action. For this scenario to work, the information that leads to gravity should be stable. The operators that are measured should therefore be very slow operators. Mathematically, this corresponds to the operators that almost commute with the Hamiltonian $\hat H$. Recently, the question of finding slow observables was addressed for one-dimensional quantum spin chains \cite{kim}. 

Further suggestions for a deep connection between gravity and measurements in quantum systems came from Penrose and Di\'osi who argued that superpositions of different spacetimes, and thus gravitational fields, should be impossible. By this argument, they were led to introduce a fundamental rate of decoherence, determined by gravitational interactions \cite{penrose,diosi}. We propose to solve the same problem in a different way. In our view, classical gravity emerges as a consequence of classical information about a quantum system and can therefore never be in a superposition. As long as the information is not available, it cannot have a gravitational effect and not influence the structure of spacetime. We agree that gravity should be restricted to the classical realm, but we disagree with Penrose on his conclusion that the ultimate theory of reality should be classical \cite{penrose14}. We opt for the quantum to be the fundamental level of description.

The idea that gravity is a thermodynamic phenomenon dates back to Jacobson, who presented a derivation of the full Einstein equations from the laws of thermodynamics, combined with the Unruh effect \cite{unruh}. We were inspired by the work by Verlinde who, building on the ideas of Jacobson, made the conjecture of an `entropic origin' of gravity \cite{verlinde}. He attributed an amount of $S_{BH}$ to each volume in space, close to our interpretation of the vacuum having high entropy.
There is however an important differences between our and Verlinde's works. Where we consider an underlying local temperature $\Theta(x)$ as the fundamental object, Verlinde takes Unruh's relation between temperature and acceleration as a postulate. Our underlying analysis allows to justify it.  

In spirit, our work is also close to the `thermal time hypothesis' of Connes and Rovelli \cite{thermaltime}. From formal mathematical considerations, they came to the conclusion that the entanglement Hamiltonian should dictate the speed of time. In subsequent work, they also connected the thermal time hypothesis to the Tolman effect by comparing the `thermal time flow' to the `mechanical time flow' \cite{smerlak,haggard}. 
We would interpret the latter as the time flow of the real Hamiltonian $\hat H$, where the former is the time we extract from the local temperature.

Gravity as an effective theory for condensed matter systems is a subject that has attracted significant interest, starting with Unruh's insight that the propagation of sound waves in a flowing fluid can be described by an effective metric \cite{unruh_eff}. When going to quantum fluids, this leads to an effective description in terms of a quantum field theory on a curved space time \cite{barcelo2005,volovik}. For example, it in has been predicted that Hawking radiation should be emitted from sonic horizons, boundaries between subsonic and supersonic flow \cite{unruh_eff}. In these types of emergent gravity, the metric is induced by the superfluid flow, which is assumed to be imposed from the outside. 

Despite the similarity with the hydrodynamic works on analog gravity in condensed matter systems, the mechanism that we propose is quite different. First, in our picture, it is not the hydrodynamic flow that generates the metric, but the local information about the quantum system. Secondly, we use a different principle for our analysis. Where the hydrodynamic gravity is based on the analysis of waves, we base our analysis on the statistics of the excitations. 

Further in the context of emergence in condensed matter systems, it was shown by d'Alessio and Polkovnikov how inertia (mass) emerges in adiabatic perturbation theory beyond the Born-Oppenheimer approximation \cite{dalessio2014}. It will be of great interest to understand the relation of their work with our view on the equivalence principle.

The first theoretical works on the entanglement entropy in extended quantum systems \cite{bombelli,srednicki,hooft1985,callan1994} were motivated by the area law for black holes \cite{bekenstein,hawking}. This research spurred the study of entanglement in condensed matter physics which has led to a great advancement of our understanding of correlated quantum systems \cite{Eisert,AmicoRMP}, but to the best of our knowledge, there is not yet a consensus on the connection to the Bekenstein-Hawking entropy.

More recently, it was suggested by Van Raamsdonk that spacetime is built up by entanglement \cite{vanraamsdonk,nv2015}. He argues that regions that disentangling parts of  a system results in the spacetime pulling apart. His arguments are based on holography, but are in spirit close to ours. When two regions are not entangled, the mutual information is zero and they cannot influence each other gravitationally. 

Van Raamsdonk's work, as well as Verlinde's together the large majority of research activity on quantum gravity is nowadays carried out in the context of adS/CFT correspondence \cite{maldacena}. The connection \cite{schwingle} between the multiscale entanglement renormalisation ansatz (MERA)  \cite{mera} for correlated quantum systems and the adS/CFT correspondence may provide some links with our approach. Also the entanglement temperature of subregions was studied in this framework \cite{casini2011,bhattacharya2013,wong2013,bianchi2014,kumar2015}. In particular, in Ref. \cite{bhattacharya2013}, a force in the direction of lower entanglement entropy was found, in agreement with our arguments of a force toward lower effective temperature.

Connections between cosmology, black holes and information theory have also been made by Lloyd \cite{lloyd2000,lloyd2002}. He bases his analyses on the computational interpretation of entropy (bits), temperature (operations per bit per unit of time) and energy (total number of operations per unit of time). Similar relations were used by Ng, in the context of space-time foam holographic models \cite{ng2008}, a picture suggested by Wheeler.

Wheeler's disciples also laid the foundation of the modern consensus on the interpretation of quantum mechanics, a hybrid of many worlds and decoherence. Everett was the first to make explicit the conclusion that follows inevitably from the linearity of quantum mechanics: in the wave function, all possibilities happen simultaneously  \cite{everett1957}. Only when measurements are made, a particular `realisation' of the world appears.
Zurek's decoherence analysis \cite{zurek2003} most clearly shows why these different realisations do not interfere (in the technical and colloquial sense) with each other. He introduced `pointer states', the states that are robust under interaction with the environment. He dubbed the fact that those pointer states are selected by the environment `einselection' (environment induced selection). These concepts are crucial in our understanding of the emergence of quantum gravity from quantum mechanics. Our picture of matter as a robust local nonequilibrium implies that all matter excitations should be pointer states, with as the ultimate pointer state a black hole.

Finally, our work is in a sense also an implementation of Wheeler's `it from bit' \cite{wheeler1996}, since we argued that the most important characteristic of matter is information, a local reduction of the entropy. 


\section{Conclusions and outlook}

In the present work, we have suggested a mechanism for time dilation that emerges out of a quantum system through a statistical analysis. It is clear that a large effort will be needed before our suggestion can become a viable candidate for a description of real quantum gravity. 

As we have discussed, information should be extracted from the quantum system through slow observables, such that the obtained information is stable. Technically, finding these observables is a difficult problem: the search takes place in an exponentially large Hilbert space. 

A further issue concerns the range the gravitational potential. In the simple model that we investigated, we found a power law decay for the mutual information in one-dimensional noninteracting fermions.  An obvious question is to find the conditions for which one obtains a $1/r$ gravitational potential at large distances in three dimensional systems.

So far, we have only provided arguments for one aspect of Einstein gravity to emerge from a statistical analysis of quantum systems, namely time dilation in a static situation. If gravity is really emerging according to our mechanism, one should be able to derive the full set of geodesic and Einstein equations from statistical considerations, with far richer phenomenology than Newtonian gravity. It should ultimately include cosmology from the perspective of closed quantum systems.

In conclusion, we conjecture that gravity appears as a deformation of statistics due to local information. This type of locally deformed quantum states are usually not considered in calculations on many body quantum systems, that are concerned with small numbers of excitations. 
It would actually seem implausible that large fluctuations in entangled quantum systems do not influence their surroundings. Two possibilities then remain if quantum mechanics is a fundamental theory. Either this influence is gravity or it is another, unknown effect. Our arguments are encouraging for the gravity scenario to be correct. The description of large fluctuations will be mathematically challenging, but a thorough understanding seems unavoidable if we want to grasp how the physical world is related to our quantum models.

\section{Acknowledgements}

We thank Jacques Tempere and Jozef Devreese for encouraging discussions. D.S. acknowledges support of the FWO as post-doctoral fellow of the Research Foundation - Flanders.

\bibliography{refs}

\begin{thebibliography}{71}%
\makeatletter
\providecommand \@ifxundefined [1]{%
 \@ifx{#1\undefined}
}%
\providecommand \@ifnum [1]{%
 \ifnum #1\expandafter \@firstoftwo
 \else \expandafter \@secondoftwo
 \fi
}%
\providecommand \@ifx [1]{%
 \ifx #1\expandafter \@firstoftwo
 \else \expandafter \@secondoftwo
 \fi
}%
\providecommand \natexlab [1]{#1}%
\providecommand \enquote  [1]{``#1''}%
\providecommand \bibnamefont  [1]{#1}%
\providecommand \bibfnamefont [1]{#1}%
\providecommand \citenamefont [1]{#1}%
\providecommand \href@noop [0]{\@secondoftwo}%
\providecommand \href [0]{\begingroup \@sanitize@url \@href}%
\providecommand \@href[1]{\@@startlink{#1}\@@href}%
\providecommand \@@href[1]{\endgroup#1\@@endlink}%
\providecommand \@sanitize@url [0]{\catcode `\\12\catcode `\$12\catcode
  `\&12\catcode `\#12\catcode `\^12\catcode `\_12\catcode `\%12\relax}%
\providecommand \@@startlink[1]{}%
\providecommand \@@endlink[0]{}%
\providecommand \url  [0]{\begingroup\@sanitize@url \@url }%
\providecommand \@url [1]{\endgroup\@href {#1}{\urlprefix }}%
\providecommand \urlprefix  [0]{URL }%
\providecommand \Eprint [0]{\href }%
\providecommand \doibase [0]{http://dx.doi.org/}%
\providecommand \selectlanguage [0]{\@gobble}%
\providecommand \bibinfo  [0]{\@secondoftwo}%
\providecommand \bibfield  [0]{\@secondoftwo}%
\providecommand \translation [1]{[#1]}%
\providecommand \BibitemOpen [0]{}%
\providecommand \bibitemStop [0]{}%
\providecommand \bibitemNoStop [0]{.\EOS\space}%
\providecommand \EOS [0]{\spacefactor3000\relax}%
\providecommand \BibitemShut  [1]{\csname bibitem#1\endcsname}%
\let\auto@bib@innerbib\@empty
\bibitem [{\citenamefont {Penrose}(2014)}]{penrose14}%
  \BibitemOpen
  \bibfield  {author} {\bibinfo {author} {\bibfnamefont {R.}~\bibnamefont
  {Penrose}},\ }\href {\doibase 10.1007/s10701-013-9770-0} {\bibfield
  {journal} {\bibinfo  {journal} {Foundations of Physics}\ }\textbf {\bibinfo
  {volume} {44}},\ \bibinfo {pages} {557} (\bibinfo {year} {2014})}\BibitemShut
  {NoStop}%
\bibitem [{\citenamefont {Aguirre}\ and\ \citenamefont
  {Tegmark}(2011)}]{tegmark}%
  \BibitemOpen
  \bibfield  {author} {\bibinfo {author} {\bibfnamefont {A.}~\bibnamefont
  {Aguirre}}\ and\ \bibinfo {author} {\bibfnamefont {M.}~\bibnamefont
  {Tegmark}},\ }\href {\doibase 10.1103/PhysRevD.84.105002} {\bibfield
  {journal} {\bibinfo  {journal} {Phys. Rev. D}\ }\textbf {\bibinfo {volume}
  {84}},\ \bibinfo {pages} {105002} (\bibinfo {year} {2011})}\BibitemShut
  {NoStop}%
\bibitem [{\citenamefont {Joos}\ \emph {et~al.}(2013)\citenamefont {Joos},
  \citenamefont {Zeh}, \citenamefont {Kiefer}, \citenamefont {Giulini},
  \citenamefont {Kupsch},\ and\ \citenamefont {Stamatescu}}]{joos2013}%
  \BibitemOpen
  \bibfield  {author} {\bibinfo {author} {\bibfnamefont {E.}~\bibnamefont
  {Joos}}, \bibinfo {author} {\bibfnamefont {H.~D.}\ \bibnamefont {Zeh}},
  \bibinfo {author} {\bibfnamefont {C.}~\bibnamefont {Kiefer}}, \bibinfo
  {author} {\bibfnamefont {D.~J.}\ \bibnamefont {Giulini}}, \bibinfo {author}
  {\bibfnamefont {J.}~\bibnamefont {Kupsch}}, \ and\ \bibinfo {author}
  {\bibfnamefont {I.-O.}\ \bibnamefont {Stamatescu}},\ }\href@noop {} {\emph
  {\bibinfo {title} {Decoherence and the appearance of a classical world in
  quantum theory}}}\ (\bibinfo  {publisher} {Springer Science \& Business
  Media},\ \bibinfo {year} {2013})\BibitemShut {NoStop}%
\bibitem [{\citenamefont {Gell-Mann}\ and\ \citenamefont
  {Hartle}(1993)}]{gell-mann1993}%
  \BibitemOpen
  \bibfield  {author} {\bibinfo {author} {\bibfnamefont {M.}~\bibnamefont
  {Gell-Mann}}\ and\ \bibinfo {author} {\bibfnamefont {J.~B.}\ \bibnamefont
  {Hartle}},\ }\href {\doibase 10.1103/PhysRevD.47.3345} {\bibfield  {journal}
  {\bibinfo  {journal} {Phys. Rev. D}\ }\textbf {\bibinfo {volume} {47}},\
  \bibinfo {pages} {3345} (\bibinfo {year} {1993})}\BibitemShut {NoStop}%
\bibitem [{\citenamefont {Hu}(2014)}]{hu2014}%
  \BibitemOpen
  \bibfield  {author} {\bibinfo {author} {\bibfnamefont {B.~L.}\ \bibnamefont
  {Hu}},\ }\href {http://stacks.iop.org/1742-6596/504/i=1/a=012021} {\bibfield
  {journal} {\bibinfo  {journal} {Journal of Physics: Conference Series}\
  }\textbf {\bibinfo {volume} {504}},\ \bibinfo {pages} {012021} (\bibinfo
  {year} {2014})}\BibitemShut {NoStop}%
\bibitem [{\citenamefont {Zubarev}\ \emph {et~al.}(1996)\citenamefont
  {Zubarev}, \citenamefont {Mozorov},\ and\ \citenamefont {R\"opke}}]{zubarev}%
  \BibitemOpen
  \bibfield  {author} {\bibinfo {author} {\bibfnamefont {D.}~\bibnamefont
  {Zubarev}}, \bibinfo {author} {\bibfnamefont {V.}~\bibnamefont {Mozorov}}, \
  and\ \bibinfo {author} {\bibfnamefont {G.}~\bibnamefont {R\"opke}},\
  }\href@noop {} {\emph {\bibinfo {title} {Statistical Mechanics of
  Nonequilibrium Processes. Volume 1: Basic Concepts, Kinetic Theory}}}\
  (\bibinfo  {publisher} {Akadamie Verlag, Berlin},\ \bibinfo {year}
  {1996})\BibitemShut {NoStop}%
\bibitem [{\citenamefont {Grandy}(2012)}]{grandy2012}%
  \BibitemOpen
  \bibfield  {author} {\bibinfo {author} {\bibfnamefont {W.}~\bibnamefont
  {Grandy}},\ }\href {https://books.google.be/books?id=UXuOuQAACAAJ} {\emph
  {\bibinfo {title} {Entropy and the Time Evolution of Macroscopic Systems}}},\
  International Series of Monographs on Physics\ (\bibinfo  {publisher} {OUP
  Oxford},\ \bibinfo {year} {2012})\BibitemShut {NoStop}%
\bibitem [{\citenamefont {Polchinski}(1998)}]{polchinski}%
  \BibitemOpen
  \bibfield  {author} {\bibinfo {author} {\bibfnamefont {J.}~\bibnamefont
  {Polchinski}},\ }\href {http://dx.doi.org/10.1017/CBO9780511618123} {\emph
  {\bibinfo {title} {String Theory}}},\ Vol.~\bibinfo {volume} {2}\ (\bibinfo
  {publisher} {Cambridge University Press},\ \bibinfo {year} {1998})\ \bibinfo
  {note} {cambridge Books Online}\BibitemShut {NoStop}%
\bibitem [{\citenamefont {Rovelli}(2004)}]{rovelli}%
  \BibitemOpen
  \bibfield  {author} {\bibinfo {author} {\bibfnamefont {C.}~\bibnamefont
  {Rovelli}},\ }\href {http://dx.doi.org/10.1017/CBO9780511755804} {\emph
  {\bibinfo {title} {Quantum Gravity}}}\ (\bibinfo  {publisher} {Cambridge
  University Press},\ \bibinfo {year} {2004})\ \bibinfo {note} {cambridge Books
  Online}\BibitemShut {NoStop}%
\bibitem [{\citenamefont {Jacobson}(1995)}]{jacobson1995}%
  \BibitemOpen
  \bibfield  {author} {\bibinfo {author} {\bibfnamefont {T.}~\bibnamefont
  {Jacobson}},\ }\href@noop {} {\bibfield  {journal} {\bibinfo  {journal}
  {Physical Review Letters}\ }\textbf {\bibinfo {volume} {75}},\ \bibinfo
  {pages} {1260} (\bibinfo {year} {1995})}\BibitemShut {NoStop}%
\bibitem [{\citenamefont {Padmanabhan}(2005)}]{Padmanabhan2005}%
  \BibitemOpen
  \bibfield  {author} {\bibinfo {author} {\bibfnamefont {T.}~\bibnamefont
  {Padmanabhan}},\ }\href {\doibase
  http://dx.doi.org/10.1016/j.physrep.2004.10.003} {\bibfield  {journal}
  {\bibinfo  {journal} {Physics Reports}\ }\textbf {\bibinfo {volume} {406}},\
  \bibinfo {pages} {49 } (\bibinfo {year} {2005})}\BibitemShut {NoStop}%
\bibitem [{\citenamefont {Verlinde}(2011{\natexlab{a}})}]{verlinde2011}%
  \BibitemOpen
  \bibfield  {author} {\bibinfo {author} {\bibfnamefont {E.}~\bibnamefont
  {Verlinde}},\ }\href@noop {} {\bibfield  {journal} {\bibinfo  {journal}
  {Journal of High Energy Physics}\ }\textbf {\bibinfo {volume} {2011}},\
  \bibinfo {pages} {1} (\bibinfo {year} {2011}{\natexlab{a}})}\BibitemShut
  {NoStop}%
\bibitem [{\citenamefont {Zurek}(2003{\natexlab{a}})}]{zurek2003}%
  \BibitemOpen
  \bibfield  {author} {\bibinfo {author} {\bibfnamefont {W.~H.}\ \bibnamefont
  {Zurek}},\ }\href {\doibase 10.1103/RevModPhys.75.715} {\bibfield  {journal}
  {\bibinfo  {journal} {Rev. Mod. Phys.}\ }\textbf {\bibinfo {volume} {75}},\
  \bibinfo {pages} {715} (\bibinfo {year} {2003}{\natexlab{a}})}\BibitemShut
  {NoStop}%
\bibitem [{\citenamefont {Wigner}\ and\ \citenamefont
  {Marx}(1995)}]{wigner1995}%
  \BibitemOpen
  \bibfield  {author} {\bibinfo {author} {\bibfnamefont {E.~P.}\ \bibnamefont
  {Wigner}}\ and\ \bibinfo {author} {\bibfnamefont {G.}~\bibnamefont {Marx}},\
  }\href@noop {} {\bibfield  {journal} {\bibinfo  {journal} {Acta Physica
  Hungarica New Series Heavy Ion Physics}\ }\textbf {\bibinfo {volume} {1}},\
  \bibinfo {pages} {87} (\bibinfo {year} {1995})}\BibitemShut {NoStop}%
\bibitem [{\citenamefont {Schlosshauer}(2007)}]{schlosshauer2007}%
  \BibitemOpen
  \bibfield  {author} {\bibinfo {author} {\bibfnamefont {M.~A.}\ \bibnamefont
  {Schlosshauer}},\ }\href@noop {} {\emph {\bibinfo {title} {Decoherence: and
  the quantum-to-classical transition}}}\ (\bibinfo  {publisher} {Springer
  Science \& Business Media},\ \bibinfo {year} {2007})\BibitemShut {NoStop}%
\bibitem [{\citenamefont {Schr{\"o}dinger}(1944)}]{schrodinger1944}%
  \BibitemOpen
  \bibfield  {author} {\bibinfo {author} {\bibfnamefont {E.}~\bibnamefont
  {Schr{\"o}dinger}},\ }\href@noop {} {\emph {\bibinfo {title} {What is life?:
  the physical aspect of the living cell; based on lectures delivered under the
  auspices of the Inst. at Trinity College, Dublin, in Feb. 1943}}}\ (\bibinfo
  {publisher} {University Press},\ \bibinfo {year} {1944})\BibitemShut
  {NoStop}%
\bibitem [{\citenamefont {Brillouin}(1953)}]{brillouin1953}%
  \BibitemOpen
  \bibfield  {author} {\bibinfo {author} {\bibfnamefont {L.}~\bibnamefont
  {Brillouin}},\ }\href@noop {} {\bibfield  {journal} {\bibinfo  {journal}
  {Journal of Applied Physics}\ }\textbf {\bibinfo {volume} {24}},\ \bibinfo
  {pages} {1152} (\bibinfo {year} {1953})}\BibitemShut {NoStop}%
\bibitem [{\citenamefont {Wong}\ \emph {et~al.}(2013)\citenamefont {Wong},
  \citenamefont {Klich}, \citenamefont {Zayas},\ and\ \citenamefont
  {Vaman}}]{wong2013}%
  \BibitemOpen
  \bibfield  {author} {\bibinfo {author} {\bibfnamefont {G.}~\bibnamefont
  {Wong}}, \bibinfo {author} {\bibfnamefont {I.}~\bibnamefont {Klich}},
  \bibinfo {author} {\bibfnamefont {L.~A.~P.}\ \bibnamefont {Zayas}}, \ and\
  \bibinfo {author} {\bibfnamefont {D.}~\bibnamefont {Vaman}},\ }\href@noop {}
  {\bibfield  {journal} {\bibinfo  {journal} {Journal of High Energy Physics}\
  }\textbf {\bibinfo {volume} {2013}},\ \bibinfo {pages} {1} (\bibinfo {year}
  {2013})}\BibitemShut {NoStop}%
\bibitem [{\citenamefont {Kumar}\ and\ \citenamefont
  {Shankaranarayanan}(2015)}]{kumar2015}%
  \BibitemOpen
  \bibfield  {author} {\bibinfo {author} {\bibfnamefont {S.~S.}\ \bibnamefont
  {Kumar}}\ and\ \bibinfo {author} {\bibfnamefont {S.}~\bibnamefont
  {Shankaranarayanan}},\ }\href@noop {} {\bibfield  {journal} {\bibinfo
  {journal} {arXiv preprint arXiv:1504.00501}\ } (\bibinfo {year}
  {2015})}\BibitemShut {NoStop}%
\bibitem [{\citenamefont {Popescu}\ \emph {et~al.}(2006)\citenamefont
  {Popescu}, \citenamefont {Short},\ and\ \citenamefont
  {Winter}}]{popescu2006}%
  \BibitemOpen
  \bibfield  {author} {\bibinfo {author} {\bibfnamefont {S.}~\bibnamefont
  {Popescu}}, \bibinfo {author} {\bibfnamefont {A.~J.}\ \bibnamefont {Short}},
  \ and\ \bibinfo {author} {\bibfnamefont {A.}~\bibnamefont {Winter}},\
  }\href@noop {} {\bibfield  {journal} {\bibinfo  {journal} {Nature Physics}\
  }\textbf {\bibinfo {volume} {2}},\ \bibinfo {pages} {754} (\bibinfo {year}
  {2006})}\BibitemShut {NoStop}%
\bibitem [{\citenamefont {Lloyd}(2013)}]{lloyd2013}%
  \BibitemOpen
  \bibfield  {author} {\bibinfo {author} {\bibfnamefont {S.}~\bibnamefont
  {Lloyd}},\ }\href@noop {} {\bibfield  {journal} {\bibinfo  {journal} {arXiv
  preprint arXiv:1307.0378}\ } (\bibinfo {year} {2013})}\BibitemShut {NoStop}%
\bibitem [{\citenamefont {Goldstein}\ \emph {et~al.}(2006)\citenamefont
  {Goldstein}, \citenamefont {Lebowitz}, \citenamefont {Tumulka},\ and\
  \citenamefont {Zangh\`{\i}}}]{Goldstein}%
  \BibitemOpen
  \bibfield  {author} {\bibinfo {author} {\bibfnamefont {S.}~\bibnamefont
  {Goldstein}}, \bibinfo {author} {\bibfnamefont {J.~L.}\ \bibnamefont
  {Lebowitz}}, \bibinfo {author} {\bibfnamefont {R.}~\bibnamefont {Tumulka}}, \
  and\ \bibinfo {author} {\bibfnamefont {N.}~\bibnamefont {Zangh\`{\i}}},\
  }\href {\doibase 10.1103/PhysRevLett.96.050403} {\bibfield  {journal}
  {\bibinfo  {journal} {Phys. Rev. Lett.}\ }\textbf {\bibinfo {volume} {96}},\
  \bibinfo {pages} {050403} (\bibinfo {year} {2006})}\BibitemShut {NoStop}%
\bibitem [{\citenamefont {Onsager}(1949)}]{onsager}%
  \BibitemOpen
  \bibfield  {author} {\bibinfo {author} {\bibfnamefont {L.}~\bibnamefont
  {Onsager}},\ }\href {\doibase 10.1111/j.1749-6632.1949.tb27296.x} {\bibfield
  {journal} {\bibinfo  {journal} {Annals of the New York Academy of Sciences}\
  }\textbf {\bibinfo {volume} {51}},\ \bibinfo {pages} {627} (\bibinfo {year}
  {1949})}\BibitemShut {NoStop}%
\bibitem [{\citenamefont {Asakura}\ and\ \citenamefont
  {Oosawa}(1954)}]{asakura}%
  \BibitemOpen
  \bibfield  {author} {\bibinfo {author} {\bibfnamefont {S.}~\bibnamefont
  {Asakura}}\ and\ \bibinfo {author} {\bibfnamefont {F.}~\bibnamefont
  {Oosawa}},\ }\href {\doibase http://dx.doi.org/10.1063/1.1740347} {\bibfield
  {journal} {\bibinfo  {journal} {The Journal of Chemical Physics}\ }\textbf
  {\bibinfo {volume} {22}},\ \bibinfo {pages} {1255} (\bibinfo {year}
  {1954})}\BibitemShut {NoStop}%
\bibitem [{\citenamefont {Vedral}(2002)}]{vedral2002}%
  \BibitemOpen
  \bibfield  {author} {\bibinfo {author} {\bibfnamefont {V.}~\bibnamefont
  {Vedral}},\ }\href {\doibase 10.1103/RevModPhys.74.197} {\bibfield  {journal}
  {\bibinfo  {journal} {Rev. Mod. Phys.}\ }\textbf {\bibinfo {volume} {74}},\
  \bibinfo {pages} {197} (\bibinfo {year} {2002})}\BibitemShut {NoStop}%
\bibitem [{\citenamefont {Wolf}\ \emph {et~al.}(2008)\citenamefont {Wolf},
  \citenamefont {Verstraete}, \citenamefont {Hastings},\ and\ \citenamefont
  {Cirac}}]{Wolf}%
  \BibitemOpen
  \bibfield  {author} {\bibinfo {author} {\bibfnamefont {M.~M.}\ \bibnamefont
  {Wolf}}, \bibinfo {author} {\bibfnamefont {F.}~\bibnamefont {Verstraete}},
  \bibinfo {author} {\bibfnamefont {M.~B.}\ \bibnamefont {Hastings}}, \ and\
  \bibinfo {author} {\bibfnamefont {J.~I.}\ \bibnamefont {Cirac}},\ }\href
  {\doibase 10.1103/PhysRevLett.100.070502} {\bibfield  {journal} {\bibinfo
  {journal} {Phys. Rev. Lett.}\ }\textbf {\bibinfo {volume} {100}},\ \bibinfo
  {pages} {070502} (\bibinfo {year} {2008})}\BibitemShut {NoStop}%
\bibitem [{\citenamefont {Tolman}(1930)}]{tolman}%
  \BibitemOpen
  \bibfield  {author} {\bibinfo {author} {\bibfnamefont {R.~C.}\ \bibnamefont
  {Tolman}},\ }\href {\doibase 10.1103/PhysRev.35.904} {\bibfield  {journal}
  {\bibinfo  {journal} {Phys. Rev.}\ }\textbf {\bibinfo {volume} {35}},\
  \bibinfo {pages} {904} (\bibinfo {year} {1930})}\BibitemShut {NoStop}%
\bibitem [{\citenamefont {Hawking}(1974)}]{hawking}%
  \BibitemOpen
  \bibfield  {author} {\bibinfo {author} {\bibfnamefont {S.~W.}\ \bibnamefont
  {Hawking}},\ }\href@noop {} {\bibfield  {journal} {\bibinfo  {journal}
  {Nature}\ }\textbf {\bibinfo {volume} {248}},\ \bibinfo {pages} {30}
  (\bibinfo {year} {1974})}\BibitemShut {NoStop}%
\bibitem [{\citenamefont {Unruh}(1976)}]{unruh}%
  \BibitemOpen
  \bibfield  {author} {\bibinfo {author} {\bibfnamefont {W.~G.}\ \bibnamefont
  {Unruh}},\ }\href {\doibase 10.1103/PhysRevD.14.870} {\bibfield  {journal}
  {\bibinfo  {journal} {Phys. Rev. D}\ }\textbf {\bibinfo {volume} {14}},\
  \bibinfo {pages} {870} (\bibinfo {year} {1976})}\BibitemShut {NoStop}%
\bibitem [{\citenamefont {Milling}\ and\ \citenamefont
  {Kendall}(2000)}]{entropic_AFM}%
  \BibitemOpen
  \bibfield  {author} {\bibinfo {author} {\bibfnamefont {A.~J.}\ \bibnamefont
  {Milling}}\ and\ \bibinfo {author} {\bibfnamefont {K.}~\bibnamefont
  {Kendall}},\ }\href {\doibase 10.1021/la991442v} {\bibfield  {journal}
  {\bibinfo  {journal} {Langmuir}\ }\textbf {\bibinfo {volume} {16}},\ \bibinfo
  {pages} {5106} (\bibinfo {year} {2000})}\BibitemShut {NoStop}%
\bibitem [{\citenamefont {Lieb}\ and\ \citenamefont {Robinson}(1972)}]{LR}%
  \BibitemOpen
  \bibfield  {author} {\bibinfo {author} {\bibfnamefont {E.~H.}\ \bibnamefont
  {Lieb}}\ and\ \bibinfo {author} {\bibfnamefont {D.~W.}\ \bibnamefont
  {Robinson}},\ }\href {http://projecteuclid.org/euclid.cmp/1103858407}
  {\bibfield  {journal} {\bibinfo  {journal} {Comm. Math. Phys.}\ }\textbf
  {\bibinfo {volume} {28}},\ \bibinfo {pages} {251} (\bibinfo {year}
  {1972})}\BibitemShut {NoStop}%
\bibitem [{\citenamefont {Eisert}\ \emph {et~al.}(2010)\citenamefont {Eisert},
  \citenamefont {Cramer},\ and\ \citenamefont {Plenio}}]{Eisert}%
  \BibitemOpen
  \bibfield  {author} {\bibinfo {author} {\bibfnamefont {J.}~\bibnamefont
  {Eisert}}, \bibinfo {author} {\bibfnamefont {M.}~\bibnamefont {Cramer}}, \
  and\ \bibinfo {author} {\bibfnamefont {M.~B.}\ \bibnamefont {Plenio}},\
  }\href {\doibase 10.1103/RevModPhys.82.277} {\bibfield  {journal} {\bibinfo
  {journal} {Rev. Mod. Phys.}\ }\textbf {\bibinfo {volume} {82}},\ \bibinfo
  {pages} {277} (\bibinfo {year} {2010})}\BibitemShut {NoStop}%
\bibitem [{\citenamefont {Bhattacharya}\ \emph {et~al.}(2013)\citenamefont
  {Bhattacharya}, \citenamefont {Nozaki}, \citenamefont {Takayanagi},\ and\
  \citenamefont {Ugajin}}]{bhattacharya2013}%
  \BibitemOpen
  \bibfield  {author} {\bibinfo {author} {\bibfnamefont {J.}~\bibnamefont
  {Bhattacharya}}, \bibinfo {author} {\bibfnamefont {M.}~\bibnamefont
  {Nozaki}}, \bibinfo {author} {\bibfnamefont {T.}~\bibnamefont {Takayanagi}},
  \ and\ \bibinfo {author} {\bibfnamefont {T.}~\bibnamefont {Ugajin}},\
  }\href@noop {} {\bibfield  {journal} {\bibinfo  {journal} {Physical review
  letters}\ }\textbf {\bibinfo {volume} {110}},\ \bibinfo {pages} {091602}
  (\bibinfo {year} {2013})}\BibitemShut {NoStop}%
\bibitem [{\citenamefont {Zurek}(2003{\natexlab{b}})}]{zurek2003discord}%
  \BibitemOpen
  \bibfield  {author} {\bibinfo {author} {\bibfnamefont {W.~H.}\ \bibnamefont
  {Zurek}},\ }\href {\doibase 10.1103/PhysRevA.67.012320} {\bibfield  {journal}
  {\bibinfo  {journal} {Phys. Rev. A}\ }\textbf {\bibinfo {volume} {67}},\
  \bibinfo {pages} {012320} (\bibinfo {year} {2003}{\natexlab{b}})}\BibitemShut
  {NoStop}%
\bibitem [{\citenamefont {Huang}(2014)}]{huang}%
  \BibitemOpen
  \bibfield  {author} {\bibinfo {author} {\bibfnamefont {Y.}~\bibnamefont
  {Huang}},\ }\href {http://stacks.iop.org/1367-2630/16/i=3/a=033027}
  {\bibfield  {journal} {\bibinfo  {journal} {New Journal of Physics}\ }\textbf
  {\bibinfo {volume} {16}},\ \bibinfo {pages} {033027} (\bibinfo {year}
  {2014})}\BibitemShut {NoStop}%
\bibitem [{\citenamefont {Verlinde}(2011{\natexlab{b}})}]{verlinde}%
  \BibitemOpen
  \bibfield  {author} {\bibinfo {author} {\bibfnamefont {E.}~\bibnamefont
  {Verlinde}},\ }\href {\doibase 10.1007/JHEP04(2011)029} {\bibfield  {journal}
  {\bibinfo  {journal} {Journal of High Energy Physics}\ }\textbf {\bibinfo
  {volume} {2011}},\ \bibinfo {eid} {29} (\bibinfo {year}
  {2011}{\natexlab{b}})}\BibitemShut {NoStop}%
\bibitem [{\citenamefont {Wald}(1984)}]{wald}%
  \BibitemOpen
  \bibfield  {author} {\bibinfo {author} {\bibfnamefont {R.~M.}\ \bibnamefont
  {Wald}},\ }\href@noop {} {\emph {\bibinfo {title} {General Relativity}}}\
  (\bibinfo  {publisher} {The University of Chicago Press},\ \bibinfo {year}
  {1984})\BibitemShut {NoStop}%
\bibitem [{\citenamefont {Hawking}(1976)}]{hawking1976}%
  \BibitemOpen
  \bibfield  {author} {\bibinfo {author} {\bibfnamefont {S.~W.}\ \bibnamefont
  {Hawking}},\ }\href {\doibase 10.1103/PhysRevD.14.2460} {\bibfield  {journal}
  {\bibinfo  {journal} {Phys. Rev. D}\ }\textbf {\bibinfo {volume} {14}},\
  \bibinfo {pages} {2460} (\bibinfo {year} {1976})}\BibitemShut {NoStop}%
\bibitem [{\citenamefont {{Preskill}}(1993)}]{preskill}%
  \BibitemOpen
  \bibfield  {author} {\bibinfo {author} {\bibfnamefont {J.}~\bibnamefont
  {{Preskill}}},\ }in\ \href@noop {} {\emph {\bibinfo {booktitle} {Black Holes,
  Membranes, Wormholes and Superstrings}}},\ \bibinfo {editor} {edited by\
  \bibinfo {editor} {\bibfnamefont {S.}~\bibnamefont {{Kalara}}}\ and\ \bibinfo
  {editor} {\bibfnamefont {D.~V.}\ \bibnamefont {{Nanopoulos}}}}\ (\bibinfo
  {year} {1993})\ p.~\bibinfo {pages} {22},\ \Eprint
  {http://arxiv.org/abs/hep-th/9209058} {hep-th/9209058} \BibitemShut {NoStop}%
\bibitem [{\citenamefont {Volovik}(2009)}]{volovik}%
  \BibitemOpen
  \bibfield  {author} {\bibinfo {author} {\bibfnamefont {G.}~\bibnamefont
  {Volovik}},\ }\href {https://books.google.be/books?id=6uj76kFJOHEC} {\emph
  {\bibinfo {title} {The Universe in a Helium Droplet}}},\ International Series
  of Monographs on Physics\ (\bibinfo  {publisher} {OUP Oxford},\ \bibinfo
  {year} {2009})\BibitemShut {NoStop}%
\bibitem [{\citenamefont {Leff}\ and\ \citenamefont {Rex}(2002)}]{leff2002}%
  \BibitemOpen
  \bibfield  {author} {\bibinfo {author} {\bibfnamefont {H.}~\bibnamefont
  {Leff}}\ and\ \bibinfo {author} {\bibfnamefont {A.~F.}\ \bibnamefont {Rex}},\
  }\href@noop {} {\emph {\bibinfo {title} {Maxwell's Demon 2 Entropy, Classical
  and Quantum Information, Computing}}}\ (\bibinfo  {publisher} {CRC Press},\
  \bibinfo {year} {2002})\BibitemShut {NoStop}%
\bibitem [{\citenamefont {Maxwell}(2012)}]{maxwell}%
  \BibitemOpen
  \bibfield  {author} {\bibinfo {author} {\bibfnamefont {J.~C.}\ \bibnamefont
  {Maxwell}},\ }\href@noop {} {\emph {\bibinfo {title} {Theory of heat}}}\
  (\bibinfo  {publisher} {Courier Corporation},\ \bibinfo {year}
  {2012})\BibitemShut {NoStop}%
\bibitem [{\citenamefont {Bohr}(1949)}]{bohr}%
  \BibitemOpen
  \bibfield  {author} {\bibinfo {author} {\bibfnamefont {N.}~\bibnamefont
  {Bohr}},\ }in\ \href@noop {} {\emph {\bibinfo {booktitle} {The Library of
  Living Philosophers, Volume 7. Albert Einstein: Philosopher-Scientist}}},\
  \bibinfo {editor} {edited by\ \bibinfo {editor} {\bibfnamefont {P.~A.}\
  \bibnamefont {Schilpp}}}\ (\bibinfo  {publisher} {Open Court},\ \bibinfo
  {year} {1949})\ pp.\ \bibinfo {pages} {199--241}\BibitemShut {NoStop}%
\bibitem [{\citenamefont {Einstein}\ \emph {et~al.}(1935)\citenamefont
  {Einstein}, \citenamefont {Podolsky},\ and\ \citenamefont {Rosen}}]{epr}%
  \BibitemOpen
  \bibfield  {author} {\bibinfo {author} {\bibfnamefont {A.}~\bibnamefont
  {Einstein}}, \bibinfo {author} {\bibfnamefont {B.}~\bibnamefont {Podolsky}},
  \ and\ \bibinfo {author} {\bibfnamefont {N.}~\bibnamefont {Rosen}},\ }\href
  {\doibase 10.1103/PhysRev.47.777} {\bibfield  {journal} {\bibinfo  {journal}
  {Phys. Rev.}\ }\textbf {\bibinfo {volume} {47}},\ \bibinfo {pages} {777}
  (\bibinfo {year} {1935})}\BibitemShut {NoStop}%
\bibitem [{\citenamefont {Kim}\ \emph {et~al.}(2015)\citenamefont {Kim},
  \citenamefont {Ba\~nuls}, \citenamefont {Cirac}, \citenamefont {Hastings},\
  and\ \citenamefont {Huse}}]{kim}%
  \BibitemOpen
  \bibfield  {author} {\bibinfo {author} {\bibfnamefont {H.}~\bibnamefont
  {Kim}}, \bibinfo {author} {\bibfnamefont {M.~C.}\ \bibnamefont {Ba\~nuls}},
  \bibinfo {author} {\bibfnamefont {J.~I.}\ \bibnamefont {Cirac}}, \bibinfo
  {author} {\bibfnamefont {M.~B.}\ \bibnamefont {Hastings}}, \ and\ \bibinfo
  {author} {\bibfnamefont {D.~A.}\ \bibnamefont {Huse}},\ }\href {\doibase
  10.1103/PhysRevE.92.012128} {\bibfield  {journal} {\bibinfo  {journal} {Phys.
  Rev. E}\ }\textbf {\bibinfo {volume} {92}},\ \bibinfo {pages} {012128}
  (\bibinfo {year} {2015})}\BibitemShut {NoStop}%
\bibitem [{\citenamefont {Penrose}(1996)}]{penrose}%
  \BibitemOpen
  \bibfield  {author} {\bibinfo {author} {\bibfnamefont {R.}~\bibnamefont
  {Penrose}},\ }\href {\doibase 10.1007/BF02105068} {\bibfield  {journal}
  {\bibinfo  {journal} {General Relativity and Gravitation}\ }\textbf {\bibinfo
  {volume} {28}},\ \bibinfo {pages} {581} (\bibinfo {year} {1996})}\BibitemShut
  {NoStop}%
\bibitem [{\citenamefont {Di\'osi}(1984)}]{diosi}%
  \BibitemOpen
  \bibfield  {author} {\bibinfo {author} {\bibfnamefont {L.}~\bibnamefont
  {Di\'osi}},\ }\href {\doibase http://dx.doi.org/10.1016/0375-9601(84)90397-9}
  {\bibfield  {journal} {\bibinfo  {journal} {Physics Letters A}\ }\textbf
  {\bibinfo {volume} {105}},\ \bibinfo {pages} {199 } (\bibinfo {year}
  {1984})}\BibitemShut {NoStop}%
\bibitem [{\citenamefont {Connes}\ and\ \citenamefont
  {Rovelli}(1994)}]{thermaltime}%
  \BibitemOpen
  \bibfield  {author} {\bibinfo {author} {\bibfnamefont {A.}~\bibnamefont
  {Connes}}\ and\ \bibinfo {author} {\bibfnamefont {C.}~\bibnamefont
  {Rovelli}},\ }\href {http://stacks.iop.org/0264-9381/11/i=12/a=007}
  {\bibfield  {journal} {\bibinfo  {journal} {Classical and Quantum Gravity}\
  }\textbf {\bibinfo {volume} {11}},\ \bibinfo {pages} {2899} (\bibinfo {year}
  {1994})}\BibitemShut {NoStop}%
\bibitem [{\citenamefont {Rovelli}\ and\ \citenamefont
  {Smerlak}(2011)}]{smerlak}%
  \BibitemOpen
  \bibfield  {author} {\bibinfo {author} {\bibfnamefont {C.}~\bibnamefont
  {Rovelli}}\ and\ \bibinfo {author} {\bibfnamefont {M.}~\bibnamefont
  {Smerlak}},\ }\href {http://stacks.iop.org/0264-9381/28/i=7/a=075007}
  {\bibfield  {journal} {\bibinfo  {journal} {Classical and Quantum Gravity}\
  }\textbf {\bibinfo {volume} {28}},\ \bibinfo {pages} {075007} (\bibinfo
  {year} {2011})}\BibitemShut {NoStop}%
\bibitem [{\citenamefont {Haggard}\ and\ \citenamefont
  {Rovelli}(2013)}]{haggard}%
  \BibitemOpen
  \bibfield  {author} {\bibinfo {author} {\bibfnamefont {H.~M.}\ \bibnamefont
  {Haggard}}\ and\ \bibinfo {author} {\bibfnamefont {C.}~\bibnamefont
  {Rovelli}},\ }\href {\doibase 10.1103/PhysRevD.87.084001} {\bibfield
  {journal} {\bibinfo  {journal} {Phys. Rev. D}\ }\textbf {\bibinfo {volume}
  {87}},\ \bibinfo {pages} {084001} (\bibinfo {year} {2013})}\BibitemShut
  {NoStop}%
\bibitem [{\citenamefont {Unruh}(1981)}]{unruh_eff}%
  \BibitemOpen
  \bibfield  {author} {\bibinfo {author} {\bibfnamefont {W.~G.}\ \bibnamefont
  {Unruh}},\ }\href {\doibase 10.1103/PhysRevLett.46.1351} {\bibfield
  {journal} {\bibinfo  {journal} {Phys. Rev. Lett.}\ }\textbf {\bibinfo
  {volume} {46}},\ \bibinfo {pages} {1351} (\bibinfo {year}
  {1981})}\BibitemShut {NoStop}%
\bibitem [{\citenamefont {Barcel�}\ \emph {et~al.}(2005)\citenamefont
  {Barcel�}, \citenamefont {Liberati},\ and\ \citenamefont
  {Visser}}]{barcelo2005}%
  \BibitemOpen
  \bibfield  {author} {\bibinfo {author} {\bibfnamefont {C.}~\bibnamefont
  {Barcel�}}, \bibinfo {author} {\bibfnamefont {S.}~\bibnamefont {Liberati}},
  \ and\ \bibinfo {author} {\bibfnamefont {M.}~\bibnamefont {Visser}},\ }\href
  {\doibase 10.1007/lrr-2005-12} {\bibfield  {journal} {\bibinfo  {journal}
  {Living Reviews in Relativity}\ }\textbf {\bibinfo {volume} {8}} (\bibinfo
  {year} {2005}),\ 10.1007/lrr-2005-12}\BibitemShut {NoStop}%
\bibitem [{\citenamefont {D'Alessio}\ and\ \citenamefont
  {Polkovnikov}(2014)}]{dalessio2014}%
  \BibitemOpen
  \bibfield  {author} {\bibinfo {author} {\bibfnamefont {L.}~\bibnamefont
  {D'Alessio}}\ and\ \bibinfo {author} {\bibfnamefont {A.}~\bibnamefont
  {Polkovnikov}},\ }\href@noop {} {\bibfield  {journal} {\bibinfo  {journal}
  {Annals of Physics}\ }\textbf {\bibinfo {volume} {345}},\ \bibinfo {pages}
  {141} (\bibinfo {year} {2014})}\BibitemShut {NoStop}%
\bibitem [{\citenamefont {Bombelli}\ \emph {et~al.}(1986)\citenamefont
  {Bombelli}, \citenamefont {Koul}, \citenamefont {Lee},\ and\ \citenamefont
  {Sorkin}}]{bombelli}%
  \BibitemOpen
  \bibfield  {author} {\bibinfo {author} {\bibfnamefont {L.}~\bibnamefont
  {Bombelli}}, \bibinfo {author} {\bibfnamefont {R.~K.}\ \bibnamefont {Koul}},
  \bibinfo {author} {\bibfnamefont {J.}~\bibnamefont {Lee}}, \ and\ \bibinfo
  {author} {\bibfnamefont {R.~D.}\ \bibnamefont {Sorkin}},\ }\href {\doibase
  10.1103/PhysRevD.34.373} {\bibfield  {journal} {\bibinfo  {journal} {Phys.
  Rev. D}\ }\textbf {\bibinfo {volume} {34}},\ \bibinfo {pages} {373} (\bibinfo
  {year} {1986})}\BibitemShut {NoStop}%
\bibitem [{\citenamefont {Srednicki}(1993)}]{srednicki}%
  \BibitemOpen
  \bibfield  {author} {\bibinfo {author} {\bibfnamefont {M.}~\bibnamefont
  {Srednicki}},\ }\href {\doibase 10.1103/PhysRevLett.71.666} {\bibfield
  {journal} {\bibinfo  {journal} {Phys. Rev. Lett.}\ }\textbf {\bibinfo
  {volume} {71}},\ \bibinfo {pages} {666} (\bibinfo {year} {1993})}\BibitemShut
  {NoStop}%
\bibitem [{\citenamefont {'t~Hooft}(1985)}]{hooft1985}%
  \BibitemOpen
  \bibfield  {author} {\bibinfo {author} {\bibfnamefont {G.}~\bibnamefont
  {'t~Hooft}},\ }\href@noop {} {\bibfield  {journal} {\bibinfo  {journal}
  {Nuclear Physics B}\ }\textbf {\bibinfo {volume} {256}},\ \bibinfo {pages}
  {727} (\bibinfo {year} {1985})}\BibitemShut {NoStop}%
\bibitem [{\citenamefont {Callan}\ and\ \citenamefont
  {Wilczek}(1994)}]{callan1994}%
  \BibitemOpen
  \bibfield  {author} {\bibinfo {author} {\bibfnamefont {C.}~\bibnamefont
  {Callan}}\ and\ \bibinfo {author} {\bibfnamefont {F.}~\bibnamefont
  {Wilczek}},\ }\href@noop {} {\bibfield  {journal} {\bibinfo  {journal} {arXiv
  preprint hep-th/9401072}\ } (\bibinfo {year} {1994})}\BibitemShut {NoStop}%
\bibitem [{\citenamefont {Bekenstein}(1973)}]{bekenstein}%
  \BibitemOpen
  \bibfield  {author} {\bibinfo {author} {\bibfnamefont {J.~D.}\ \bibnamefont
  {Bekenstein}},\ }\href {\doibase 10.1103/PhysRevD.7.2333} {\bibfield
  {journal} {\bibinfo  {journal} {Phys. Rev. D}\ }\textbf {\bibinfo {volume}
  {7}},\ \bibinfo {pages} {2333} (\bibinfo {year} {1973})}\BibitemShut
  {NoStop}%
\bibitem [{\citenamefont {Amico}\ \emph {et~al.}(2008)\citenamefont {Amico},
  \citenamefont {Fazio}, \citenamefont {Osterloh},\ and\ \citenamefont
  {Vedral}}]{AmicoRMP}%
  \BibitemOpen
  \bibfield  {author} {\bibinfo {author} {\bibfnamefont {L.}~\bibnamefont
  {Amico}}, \bibinfo {author} {\bibfnamefont {R.}~\bibnamefont {Fazio}},
  \bibinfo {author} {\bibfnamefont {A.}~\bibnamefont {Osterloh}}, \ and\
  \bibinfo {author} {\bibfnamefont {V.}~\bibnamefont {Vedral}},\ }\href
  {\doibase 10.1103/RevModPhys.80.517} {\bibfield  {journal} {\bibinfo
  {journal} {Rev. Mod. Phys.}\ }\textbf {\bibinfo {volume} {80}},\ \bibinfo
  {pages} {517} (\bibinfo {year} {2008})}\BibitemShut {NoStop}%
\bibitem [{\citenamefont {Van~Raamsdonk}(2010)}]{vanraamsdonk}%
  \BibitemOpen
  \bibfield  {author} {\bibinfo {author} {\bibfnamefont {M.}~\bibnamefont
  {Van~Raamsdonk}},\ }\href {\doibase 10.1142/S0218271810018529} {\bibfield
  {journal} {\bibinfo  {journal} {International Journal of Modern Physics D}\
  }\textbf {\bibinfo {volume} {19}},\ \bibinfo {pages} {2429} (\bibinfo {year}
  {2010})}\BibitemShut {NoStop}%
\bibitem [{\citenamefont {Add}(2015)}]{nv2015}%
  \BibitemOpen
  \bibfield  {author} {\bibinfo {author} {\bibnamefont {Add}},\ }\href@noop {}
  {\bibfield  {journal} {\bibinfo  {journal} {Nature Physics}\ } (\bibinfo
  {year} {2015})}\BibitemShut {NoStop}%
\bibitem [{\citenamefont {Maldacena}(1999)}]{maldacena}%
  \BibitemOpen
  \bibfield  {author} {\bibinfo {author} {\bibfnamefont {J.}~\bibnamefont
  {Maldacena}},\ }\href {\doibase 10.1023/A:1026654312961} {\bibfield
  {journal} {\bibinfo  {journal} {International Journal of Theoretical
  Physics}\ }\textbf {\bibinfo {volume} {38}},\ \bibinfo {pages} {1113}
  (\bibinfo {year} {1999})}\BibitemShut {NoStop}%
\bibitem [{\citenamefont {Swingle}(2012)}]{schwingle}%
  \BibitemOpen
  \bibfield  {author} {\bibinfo {author} {\bibfnamefont {B.}~\bibnamefont
  {Swingle}},\ }\href {\doibase 10.1103/PhysRevD.86.065007} {\bibfield
  {journal} {\bibinfo  {journal} {Phys. Rev. D}\ }\textbf {\bibinfo {volume}
  {86}},\ \bibinfo {pages} {065007} (\bibinfo {year} {2012})}\BibitemShut
  {NoStop}%
\bibitem [{\citenamefont {Vidal}(2008)}]{mera}%
  \BibitemOpen
  \bibfield  {author} {\bibinfo {author} {\bibfnamefont {G.}~\bibnamefont
  {Vidal}},\ }\href {\doibase 10.1103/PhysRevLett.101.110501} {\bibfield
  {journal} {\bibinfo  {journal} {Phys. Rev. Lett.}\ }\textbf {\bibinfo
  {volume} {101}},\ \bibinfo {pages} {110501} (\bibinfo {year}
  {2008})}\BibitemShut {NoStop}%
\bibitem [{\citenamefont {Casini}\ \emph {et~al.}(2011)\citenamefont {Casini},
  \citenamefont {Huerta},\ and\ \citenamefont {Myers}}]{casini2011}%
  \BibitemOpen
  \bibfield  {author} {\bibinfo {author} {\bibfnamefont {H.}~\bibnamefont
  {Casini}}, \bibinfo {author} {\bibfnamefont {M.}~\bibnamefont {Huerta}}, \
  and\ \bibinfo {author} {\bibfnamefont {R.~C.}\ \bibnamefont {Myers}},\
  }\href@noop {} {\bibfield  {journal} {\bibinfo  {journal} {Journal of High
  Energy Physics}\ }\textbf {\bibinfo {volume} {2011}},\ \bibinfo {pages} {1}
  (\bibinfo {year} {2011})}\BibitemShut {NoStop}%
\bibitem [{\citenamefont {Bianchi}\ and\ \citenamefont
  {Myers}(2014)}]{bianchi2014}%
  \BibitemOpen
  \bibfield  {author} {\bibinfo {author} {\bibfnamefont {E.}~\bibnamefont
  {Bianchi}}\ and\ \bibinfo {author} {\bibfnamefont {R.~C.}\ \bibnamefont
  {Myers}},\ }\href {http://stacks.iop.org/0264-9381/31/i=21/a=214002}
  {\bibfield  {journal} {\bibinfo  {journal} {Classical and Quantum Gravity}\
  }\textbf {\bibinfo {volume} {31}},\ \bibinfo {pages} {214002} (\bibinfo
  {year} {2014})}\BibitemShut {NoStop}%
\bibitem [{\citenamefont {Lloyd}(2000)}]{lloyd2000}%
  \BibitemOpen
  \bibfield  {author} {\bibinfo {author} {\bibfnamefont {S.}~\bibnamefont
  {Lloyd}},\ }\href@noop {} {\bibfield  {journal} {\bibinfo  {journal}
  {Nature}\ }\textbf {\bibinfo {volume} {406}},\ \bibinfo {pages} {1047}
  (\bibinfo {year} {2000})}\BibitemShut {NoStop}%
\bibitem [{\citenamefont {Lloyd}(2002)}]{lloyd2002}%
  \BibitemOpen
  \bibfield  {author} {\bibinfo {author} {\bibfnamefont {S.}~\bibnamefont
  {Lloyd}},\ }\href {\doibase 10.1103/PhysRevLett.88.237901} {\bibfield
  {journal} {\bibinfo  {journal} {Phys. Rev. Lett.}\ }\textbf {\bibinfo
  {volume} {88}},\ \bibinfo {pages} {237901} (\bibinfo {year}
  {2002})}\BibitemShut {NoStop}%
\bibitem [{\citenamefont {Ng}(2008)}]{ng2008}%
  \BibitemOpen
  \bibfield  {author} {\bibinfo {author} {\bibfnamefont {Y.~J.}\ \bibnamefont
  {Ng}},\ }\href@noop {} {\bibfield  {journal} {\bibinfo  {journal} {Entropy}\
  }\textbf {\bibinfo {volume} {10}},\ \bibinfo {pages} {441} (\bibinfo {year}
  {2008})}\BibitemShut {NoStop}%
\bibitem [{\citenamefont {Everett}(1957)}]{everett1957}%
  \BibitemOpen
  \bibfield  {author} {\bibinfo {author} {\bibfnamefont {H.}~\bibnamefont
  {Everett}},\ }\href {\doibase 10.1103/RevModPhys.29.454} {\bibfield
  {journal} {\bibinfo  {journal} {Rev. Mod. Phys.}\ }\textbf {\bibinfo {volume}
  {29}},\ \bibinfo {pages} {454} (\bibinfo {year} {1957})}\BibitemShut
  {NoStop}%
\bibitem [{\citenamefont {Wheeler}(1996)}]{wheeler1996}%
  \BibitemOpen
  \bibfield  {author} {\bibinfo {author} {\bibfnamefont {J.~A.}\ \bibnamefont
  {Wheeler}},\ }\href@noop {} {\bibfield  {journal} {\bibinfo  {journal} {At
  Home in the Universe, 300 pp.. AIP Press. Also Masters of Modern Physics}\
  }\textbf {\bibinfo {volume} {1}} (\bibinfo {year} {1996})}\BibitemShut
  {NoStop}%
\end{thebibliography}%

\appendix

\section{Free fermion systems}

A Hamiltonian of the type
\begin{equation}
\hat H = \sum_{i,j} \hat a^\dag_i h_{ij} \hat a_j
\end{equation}
can be diagonalized by a unitary transformation to new operators $\hat \alpha_k$ 
\begin{equation}
\hat a_i = \sum_{k} U_{ik} \hat \alpha_k,
\end{equation}
that renders the Hamiltonian diagonal:
\begin{equation}
\hat H = \sum_{k} \epsilon_k \alpha^\dag_k  \hat \alpha_k.
\end{equation}

A density matrix of the form 
\begin{equation}
\hat \rho = e^{-\hat H}
\label{eq:aprho}
\end{equation}
has correlation functions that are diagonal in the $\hat \alpha_k$ operators
\begin{equation}
\langle \hat \alpha^\dag_k \hat \alpha_k \rangle = p_k = \frac{1}{e^{\epsilon_k} +1}.
\end{equation}
The correlation functions in terms of the old operators read
\begin{equation}
\langle a_i^\dag a_j \rangle = \left(  U P U^\dag \right)_{ij},
\end{equation}
where $P_{kl}=p_k \delta_{k,l}$.
The unitary transformation $U$ can thus be determined by diagonalising the correlation matrix $\langle a_i^\dag a_j \rangle $.
The von Neumann entropy of $\hat \rho$ can be computed as
\begin{equation}
S(\hat \rho) = - \sum_k \left[ n_k \ln n_k + (1-n_k)\ln(1-n_k)   \right].
\end{equation}

The Schmidt decomposition of a pure state
\begin{equation}
\vert \psi \rangle = \sum_i c_i | \psi_i \rangle^{(A)} | \psi_i \rangle^{(S\setminus A)} 
\end{equation}
can also be straightforwardly determined. The states $| \psi_i \rangle^{(A)}$ correspond to the eigenstates of the reduced density matrix $\rho_A$, that can be written in the form \eqref{eq:aprho}.
Their correlation function is 
\begin{equation}
\langle \hat \alpha^\dag_k \hat \alpha_k \rangle = n_k,
\end{equation}
where $n_k$ is either zero or one, since they have to be pure states.
The probability for a state with a given sequence $n_k$ to occur is given by 
\begin{equation}
{\rm prob}(\{n_k \}) = \prod_k \pi(n_k,p_k).
\end{equation}
Here $\pi(1,p_k)=p_k$ and $\pi(0,p_k)=1-p_k$ is the probability that a level is full or empty respectively. This allows to generate the Schmidt components and to compute the conditional entropy and the energy fluctuations shown in Fig. \ref{fig:IAB}.  

\end{document}